\documentclass[9pt,conference]{IEEEtran}
\IEEEoverridecommandlockouts

\usepackage{amsmath,amssymb,amsfonts}
\usepackage{graphicx}
\usepackage{svg}
\usepackage{textcomp}
\usepackage{xcolor}
\usepackage{multirow}
\usepackage{textgreek}
\usepackage{algorithm}
\usepackage{algpseudocode}
\usepackage{multirow}
\usepackage{algorithm}
\usepackage{color}

\usepackage{enumitem}
\usepackage{amssymb}
\usepackage{pifont}

\usepackage{cite}

\usepackage{textcomp}
\def\BibTeX{{\rm B\kern-.05em{\sc i\kern-.025em b}\kern-.08em
    T\kern-.1667em\lower.7ex\hbox{E}\kern-.125emX}}

\usepackage{amsthm}

\usepackage{graphicx}

\usepackage{fontawesome5} 

\usepackage{multicol,multirow}
\usepackage{array}
\usepackage{subfigure}

\definecolor{myblue}{RGB}{220,230,245}

\usepackage[markup=default]{changes} 

\usepackage{tikz}

\definecolor{gold}{RGB}{255, 215, 0}

\usetikzlibrary{shapes.geometric, calc}

\newcommand\scorebw[2]{%
  \pgfmathsetmacro\pgfxa{#1 + 1}%
  \tikzstyle{scorestars}=[star, star points=5, star point ratio=2.25, draw=gray!60, inner sep=0.15em, anchor=outer point 3]%
  \begin{tikzpicture}[baseline, scale=1]
    \foreach \i in {1, ..., #2} {
      \pgfmathparse{\i<=#1 ? "gold" : "white"}
      \edef\starcolor{\pgfmathresult}
      \draw (\i*0.85em, 0) node[name=star\i, scorestars, fill=\starcolor]  {};
    }
    \pgfmathparse{#1>int(#1) ? int(#1+1) : 0}
    \let\partstar=\pgfmathresult
    \ifnum\partstar>0
      \pgfmathsetmacro\starpart{#1-(int(#1)}
      \path [clip] ($(star\partstar.outer point 3)!(star\partstar.outer point 2)!(star\partstar.outer point 4)$) rectangle 
      ($(star\partstar.outer point 2 |- star\partstar.outer point 1)!\starpart!(star\partstar.outer point 1 -| star\partstar.outer point 5)$);
      \fill (\partstar*0.85em, 0) node[scorestars, fill=gold]  {};
    \fi
  \end{tikzpicture}%
}

\definechangesauthor[name={runzhi}, color=blue]{runzhi}

\begin{document}
\title{
Lorecast: \underline{L}ayout-Aware Performance and \\ Power F\underline{orecast}ing from Natural Language 
\thanks{}
}

\author{\IEEEauthorblockN{Runzhi Wang\IEEEauthorrefmark{1},Prianka Sengupta\IEEEauthorrefmark{1},Cristhian Roman-Vicharra\IEEEauthorrefmark{1},Yiran Chen\IEEEauthorrefmark{2}, Jiang Hu\IEEEauthorrefmark{1}}
\IEEEauthorblockA{\IEEEauthorrefmark{1}Texas A\&M University}
\IEEEauthorblockA{\IEEEauthorrefmark{2}Duke University } 
\vspace{-25pt}
}

\maketitle
\begin{abstract}
In chip design planning, obtaining reliable performance and power forecasts for various design options is of critical importance. Traditionally, this involves using system-level models, which often lack accuracy, or trial synthesis, which is both labor-intensive and time-consuming.  We introduce a new methodology, called Lorecast, which accepts English prompts as input to rapidly generate layout-aware performance and power estimates. This approach bypasses the need for HDL code development and synthesis, making it both fast and user-friendly. Experimental results demonstrate that Lorecast achieves accuracy within a few percent of error compared to post-layout analysis, while significantly reducing turnaround time.

\end{abstract}


\thispagestyle{empty}
\section{Introduction}
\label{sec:Introduction}

%
Chip design planning is the process of evaluating and optimizing key metrics such as performance and power during the early stages of hardware development. In this process, predicting the performance and power of various design options is a critical yet challenging task. Engineers rely on these performance metrics to make informed decisions about architectural choices, resource allocation, and overall design optimization. 
For instance, designers might pose questions such as, ``If the unfolding factor for an Infinite Impulse Response (IIR) filter is adjusted from $x$ to $y$, will the block-level power constraint be violated?" or ``Will replacing a carry-ripple adder with a carry-lookahead adder create significant challenges for timing closure?"

Although system-level models and transaction-level models~\cite{black2004systemc} are useful in planning, they are loosely timed or approximately timed and thus incapable of providing sufficiently accurate estimates. Architecture-level models, such as McPAT~\cite{li2009mcpat}, are mostly restricted to microprocessor designs and can have very large errors. 
For example, the power estimate error can be as large as $200\%$~\cite{xi2015quantifying}.
Errors at early design stages can propagate throughout the design process, resulting in costly revisions, delays, or suboptimal designs. In competitive markets such as Artificial Intelligence (AI) accelerators, Internet of Things (IoT), and high-performance computing, time-to-market is crucial, and slow iterations can hinder innovation.

Alternatively, designers can write Hardware Description Language (HDL) code, followed by logic and layout synthesis, which demands significant time and labor investments. The accuracy of architecture-level models can also be improved by machine learning-based calibration~\cite{zhai2021mcpat}. However, the calibration still requires expensive Register Transfer Level (RTL) implementation and synthesis. 
While some attempts have been made to predict performance and power using designer-written HDL code, creating the HDL code itself is a time-consuming and challenging process. When a designer has an idea, it often takes several times longer to translate that idea into HDL code.
This creates a bottleneck, especially in the early design stages, where rapid iterations are needed to efficiently explore the design space. The challenge lies in providing a quick and reliable way to estimate performance and power based on a designer's idea.
\begin{figure}[!htb] 
\centering 
\vspace{-5pt}
\includegraphics[width=1.0\linewidth]{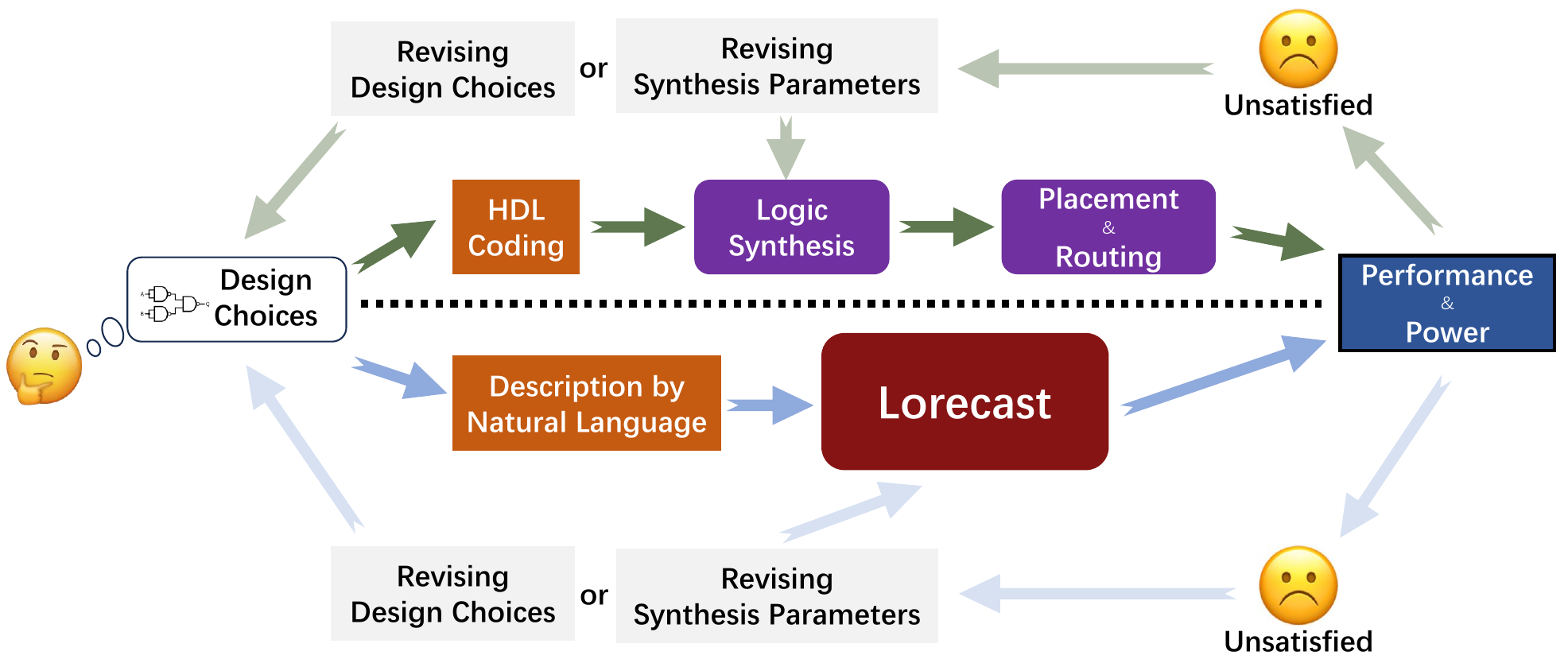}
\vspace{-3pt}
\caption{Traditional flow vs. proposed flow with Lorecast approach: from natural language to performance and power.}
\vspace{-12pt}
\label{fig:flow comparison}
\end{figure}

To address this problem, we propose an approach called Lorecast, which transforms a designer's natural language description of their idea directly into layout-aware performance and power estimates. Figure~\ref{fig:flow comparison} illustrates how Lorecast streamlines the traditional flow by avoiding manual HDL coding, logic and layout synthesis, while enabling natural-language-driven performance and power estimation in early design stages. The flow with Lorecast accelerates design planning, enabling faster and more efficient exploration of design options. Moreover, Lorecast reduces the need for system/architecture-level designers to have in-depth knowledge of HDL code. 
A key component of Lorecast is Large Language Model (LLM)-based automatic Verilog code generation. Distinguished from existing approaches, Lorecast significantly relaxes the requirement for functional correctness of the generated Verilog code. Functional correctness has been a difficult challenge to the practical and large-scale adoption of LLM-based Verilog code generation. However, Lorecast mitigates this challenge by using syntax-correct and structurally similar Verilog code generated by LLMs, which is sufficient for its purposes. The novelty of Lorecast primarily lies in its integration of LLM-based Verilog code generation with Machine Learning (ML)-based predictive models, effectively mitigating the limitations of standalone LLM-generated code. Without solving the challenge of functional correctness, the practical utility of LLM-generated Verilog code remains extremely limited, if not entirely nonexistent. Lorecast significantly broadens the applicability of such code by relaxing the requirement of functional correctness, effectively breathing life into LLM-generated Verilog designs.







The contributions of this work are summarized as follows.

\vspace*{-1mm}
\begin{itemize}
 \item \textbf{Performance and power prediction from natural language.} The Lorecast framework allows designers to obtain performance and power estimates directly from English descriptions of their ideas, bypassing the need for HDL coding. 
 \item \textbf{High forecast accuracy.} Lorecast achieves accurate estimates, with an average percentage error of only 2\% for both power and Total Negative Slack (TNS), compared to the post-layout analysis by a commercial tool. Even for cases where the Verilog codes generated by the LLM are functionally incorrect, the average forecast errors are 5\% and 7\% for power and TNS, respectively, compared to analysis of functionally correct designs.
 \item \textbf{Acceleration of the design planning process.} Lorecast accelerates the performance and power estimation process by $4.6\times$ compared to conventional methods that involve manually writing Verilog code, logic and layout synthesis. 
  \item \textbf{Relaxation of functional correctness requirement.} We provide analysis and experimental evidence demonstrating that Lorecast remains effective even when the generated Verilog code is functionally incorrect, primarily due to its structural similarity to functionally correct designs.
 \item \textbf{Enhancement of syntax correctness through prompting techniques.} We propose two complementary techniques to improve the syntax correctness of LLM-generated code. The first technique introduces pseudocode as an intermediate reasoning step to guide HDL code generation, resulting in a 33\% to 43\% improvement in syntax correctness compared to free narrative-style prompting. Building on this, the second technique employs iterative prompting with regulated feedback, improving syntax correctness by 6\% over na\"ive error-based iteration and 11\% over direct generation without feedback. 
 \item \textbf{Evaluation with circuits significantly larger than existing ones.} So far, LLM-based Verilog code generation techniques are mostly restricted to small circuits. We developed circuit cases that double the sizes in terms of cell count compared to the latest publicly released testcases.  
 On these larger cases, Lorecast achieved $100\%$ syntax correctness. 
 \item {\bf Superiority over direct LLM-based forecasting methods.} Experimental results show that Lorecast is a much more promising approach than direct LLM-based (including a fine-tuned LLM) forecasting without generating Verilog code. 
\end{itemize}

\section{Related Works}
\label{sec:PreviousWorks}

To the best of our knowledge, there is no prior study on forecasting circuit performance and power from natural language. However, there have been related works, which are briefly reviewed as follows.

\vspace*{1mm}
\noindent
{\bf Leveraging LLMs to generate HDL code.}
Previous research has explored the feasibility of using LLMs for design tasks. ChatChisel\cite{liu2024ChatChisel} leveraged LLMs alongside collaboration and Retrieval-Augmented Generation (RAG)\cite{Lewis2020RAG} techniques to enhance code generation performance, successfully producing a RISC-V CPU and demonstrating the potential of LLMs in generating complex circuits. ChatCPU\cite{Wang2024ChatCPU} was a framework combining LLMs with CPU design automation, which has been used to successfully design a CPU and complete its tape-out. BetterV\cite{Pei2024BetterV} utilized a discriminator to guide Verilog code generation. Other studies focused on using LLMs to assist in hardware design. ChipGPT\cite{Chang2023ChipGPT} demonstrated that LLMs can aid users in understanding complex designs. VGV\cite{Wong2024VGV} took advantage of LLMs’ capabilities in computer vision to generate Verilog code directly from circuit diagrams.

\vspace*{1mm}
\noindent
{\bf Evaluating Verilog code generation capabilities of LLMs.} Some studies focused on evaluating the code generation capabilities of LLMs, primarily in terms of syntax and functional correctness. In \cite{Chang2023ChipGPT}, the code generation capabilities of ChatGPT were compared to other LLMs.  VerilogEval\cite{liu2023VerilogEval} introduced a benchmark for evaluating the correctness of Verilog code generation, along with an automated evaluation framework. RTLLM\cite{lu2024RTLLM} proposed a benchmark and evaluated how prompt styles impact code generation accuracy, also presenting a prompting technique to improve correctness. RTL-Repo\cite{allam2024RTL-Repo} developed a large-scale benchmark for assessing Verilog code generation capabilities of LLMs, containing over 4,000 Verilog code samples. CreativEval\cite{DeLorenzo2024CreativEval} proposed a new perspective by focusing not on correctness but on fluency, flexibility, originality, and refinement.

\vspace*{1mm}
\noindent
{\bf Enhancing Verilog code generation capabilities of LLMs.} To improve the accuracy of code generation, some researchers have experimented with fine-tuning open-source and lightweight LLM models \cite{Thakur2023Benchmarking}\cite{Liu2024RTLCoder}\cite{thakur2024verigen}. In \cite{Nadimi2024Multi-Expert_LLM}, researchers used a framework that integrates datasets categorized by different design complexities to fine-tune LLMs for specific tasks. RTLLM\cite{lu2024RTLLM} proposed a structured prompt technique that guides LLMs in generating HDL code. Autochip \cite{Thakur2024AutoChip} proposed a framework that uses feedback from syntax-checking tools to correct syntax errors in LLM-generated code, however it still can not guarantee the syntax correctness. 
OriGen \cite{Cui2024OriGen} used feedback-based correction to collect datasets for augmentation and improve code generation.
Both RTLFixer \cite{Tsai2024RTLFixer} and AutoVCoder \cite{Gao2024AutoVCoder} utilized RAG to enhance syntax error correction in code. EDA Corpus \cite{Wu2024EDACorpus} and MG-Verilog \cite{Zhang2024MG-Verilog} proposed datasets tailored to various application scenarios to improve Verilog code generation capabilities in LLMs. Additionally, other researchers have proposed frameworks for generating datasets specifically for fine-tuning LLMs \cite{Chang2024Data_all_you_need}. VerilogReader\cite{Ma2024VerilogReader} proposed a framework to expand the comprehension scope of LLMs for improved generation of Verilog test code.

\vspace*{1mm}
\noindent
{\bf Leveraging LLMs to generate HDL testbenches.}
Recent studies have explored using LLMs to automate Verilog testbench generation. AutoBench \cite{qiu2024autobench} generated self-checking testbenches from design descriptions via scenario extraction and syntax correction. CorrectBench \cite{qiu2024correctbench} was built on AutoBench by adding automatic validation and iterative self-correction using behavior comparisons across RTL variants. 
Although CorrectBench introduced enhancements to improve functional accuracy and coverage, it ultimately inherited the same fundamental limitation as AutoBench: the generated testbenches often failed to ensure correctness.

\vspace*{1mm}
\noindent
{\bf Predicting PPA from HDL code.} Some studies focused on obtaining design evaluations during the early design stages. In \cite{Sengupta2022Prediction}, a machine learning method was proposed to perform Verilog-based evaluations without requiring synthesis. Some researchers concentrated on predictions specifically based on synthesis results \cite{Xu2022systhesisPrediction}. MasterRTL \cite{Fang2023MasterRTL} introduced an approach for design evaluation using a simple operator graph to describe relationships between logic gates. In \cite{Moravej2024LUTGraph}, researchers leveraged the Look-Up Table (LUT) Graph from Verilog code to predict. Additionally, some researchers employed Graph Neural Networks (GNNs) to estimate the design’s maximum arrival time by predicting component delays and slopes \cite{Lopera2023GNNPrediction}.


\begin{figure*}[ht]
    \centering
    \vspace{-5pt}
    \includegraphics[width=1\textwidth]{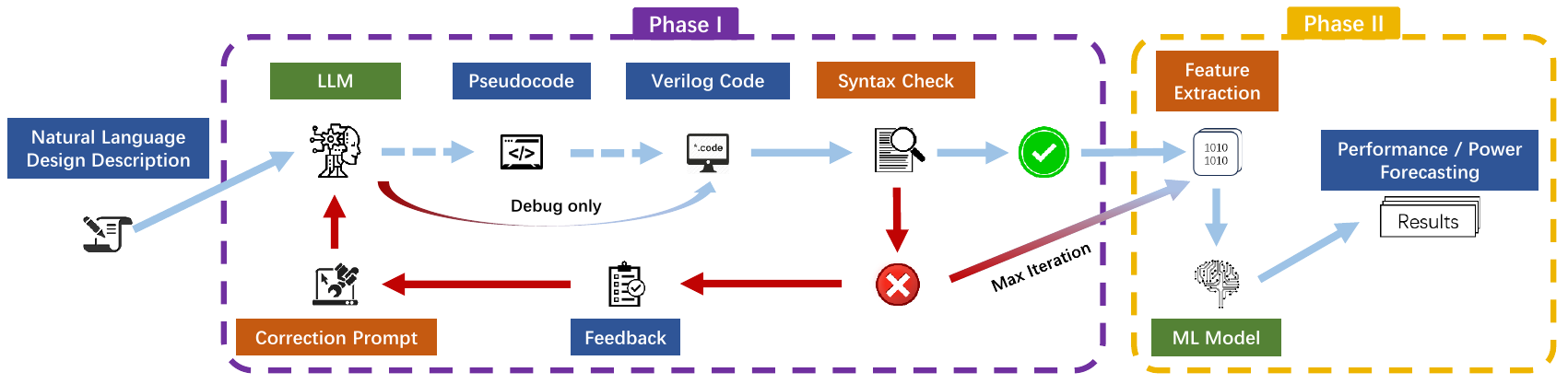}
    \vspace{-15pt}
    \caption{Overview of the Lorecast methodology.}
    \label{fig:workfolw1}
\vspace{-15pt}
\end{figure*}

\section{Background}
\label{sec:background}

\subsection{Large Language Models }
\label{sec:graph_clustering}
Large Language Models (LLMs) are advanced AI systems designed to understand and generate human-like text by processing vast amounts of language data\cite{Radford2019LLM}. These models are typically based on Transformer architectures\cite{Vaswani2017Transformer} and employ billions of parameters to capture complex patterns and relationships within language, enabling them to perform tasks such as text generation, translation, summarization, and question answering\cite{Brown2020LLMLearners}. Prominent examples include OpenAI’s GPT-4 \cite{OpenAI2023GPT-4}, Google’s Gemini 1.5 Pro \cite{Google2024Gemini}, and DeepSeek \cite{liu2024deepseek}, showcasing the immense capabilities of LLMs. With 175 billion parameters and a context capacity of 1 million tokens, these models achieve near-human performance across a variety of Natural Language Processing (NLP) tasks. The primary strength of LLMs lies in their pretraining on extensive text corpora, which gives them a general understanding of language that can be fine-tuned for specific tasks or domains \cite{Thakur2023Benchmarking}\cite{Liu2024RTLCoder}\cite{Bommasani2022finite}. As LLMs continue to evolve, their applications are expanding across various fields, including EDA, where they are increasingly employed for tasks such as code generation\cite{Wang2024ChatCPU}\cite{Pei2024BetterV}\cite{Wong2024VGV}.
\subsection{Verilog and Code Correctness}
Verilog is a hardware description language (HDL) widely used in digital design and circuit development for specifying and modeling electronic systems at various abstraction levels, from high-level functional descriptions to detailed structural representations \cite{Palnitkar2003Verilog}. Ensuring code correctness in Verilog is crucial because errors at the HDL level can lead to significant functional failures, performance inefficiencies, and increased costs when translated to physical hardware. Code correctness in Verilog includes both syntax correctness, ensuring code is free from syntax errors, and functional correctness, validating that the code’s behavior aligns with design specifications \cite{Harris2010CMOS}. While recent developments\cite{Wang2024ChatCPU}\cite{Pei2024BetterV} in LLM-based code generation have sought to improve Verilog code correctness via automated synthesis and error detection, their effectiveness remains limited—correctness rates often remain low even after such enhancements, due to the lack of deeper semantic understanding, global planning, generalizable control and checking mechanisms.

\section{The Proposed Lorecast Methodology}
\label{sec:Methods}

\subsection{Overview}

The goal of Lorecast is to take natural language prompts as input and produce layout-aware performance and power forecasts for the circuit corresponding to the prompts. An overview of the Lorecast methodology is provided in Figure~\ref{fig:workfolw1}. 
 It consists of two phases. 
 Phase I is LLM-based Verilog code generation, and Phase II is performance/power forecasting according to the generated Verilog code.
Although LLM-based Verilog code generation has been studied in prior works, our approach differs significantly in a crucial aspect: functional correctness is far less critical in our case. In conventional approaches~\cite{liu2023VerilogEval} \cite{lu2024RTLLM}, functional correctness is essential because the generated Verilog code is typically intended for synthesis. By contrast, in our methodology, functional errors usually have a very small impact on performance and power forecasting.
Functional correctness remains a significant challenge for LLM-based Verilog code generation and is far from being well solved. Our innovative use of LLM-based Verilog code generation largely bypasses this challenge. As such, we can focus on syntax correctness, which is much more achievable. Additionally, using Verilog code as an intermediate representation enables significantly better forecasting accuracy compared to direct performance and power predictions using LLMs.

\subsection{Phase I: LLM-Based Verilog Code Generation}
\label{sec:code_generation}


As shown in Figure~\ref{fig:workfolw1}, the LLM-based Verilog code generation primarily has two components:
(1) The LLM takes English prompts as input and produces corresponding Verilog code; (2) Syntax check is performed on the code and corrective prompts with feedback are fed to the LLM again for producing improved code.
The LLMs for Verilog code generation can be obtained from either existing closed-source models, such as ChatGPT~\cite{OpenAI2023GPT-4} and Gemini~\cite{Google2024Gemini}, or fine-tuning open-source models, such as Llama~\cite{meta2024Llama}. A recent analysis in \cite{Liu2024RTLCoder} shows that the best results so far were obtained from GPT4~\cite{OpenAI2023GPT-4}, a closed-source model. Therefore, we focus on using closed-source models with prompt engineering enhancements. Although the focus of our study here is closed-source models, our methodology is general and can work with fine-tuned open-source models as well.

\begin{table}[htb]
\vspace{-10pt}
\caption{Effect of Prompt and Feedback Design on LLM Stability and Code Generation Quality.
}
\centering
\resizebox{1\linewidth}{!}{%
\begin{tabular}{l||c|c|c||c|c}
\hline
\multicolumn{1}{c||}{Methodology} & \begin{tabular}[c]{@{}c@{}}Prompt structured\\ level\end{tabular} & \begin{tabular}[c]{@{}c@{}}Intermediate\\ representation\end{tabular} & \begin{tabular}[c]{@{}c@{}}Feedback\\ mechanism\end{tabular} & \begin{tabular}[c]{@{}c@{}}LLM\\ non-determinism\end{tabular} & \begin{tabular}[c]{@{}c@{}}Syntax\\ correctness\end{tabular} \\ \hline
\hline
Free-form & \ding{55} None & \ding{55} None & \ding{55} None & \begin{tabular}[c]{@{}c@{}}Very high\\ No constrained\end{tabular} & \scalebox{0.8}{\scorebw{1.5}{5}}\\ \hline
VerilogEval\cite{liu2023VerilogEval} & \ding{51} Basic & \ding{55} None & \ding{55} None & \begin{tabular}[c]{@{}c@{}}High\\ Partially constrained\end{tabular} & \scalebox{0.8}{\scorebw{2.5}{5}} \\ \hline
RTLLM\cite{lu2024RTLLM} & \ding{51} Highly & \ding{55} None & \ding{55} None & \begin{tabular}[c]{@{}c@{}}Moderate\\ Clearly constrained\end{tabular} & \scalebox{0.8}{\scorebw{3.9}{5}} \\ \hline
Autochip\cite{Thakur2024AutoChip} & \ding{51} Basic & \ding{55} None & \ding{51} Unstructured & \begin{tabular}[c]{@{}c@{}}Moderate\\ Errors can be fixed\end{tabular} & \scalebox{0.8}{\scorebw{3.7}{5}} \\ \hline
Lorecast & \ding{51} Highly & \ding{51} Pseudocode & \ding{51} Structured & \begin{tabular}[c]{@{}c@{}}Very low\\ Fully constrained\end{tabular} & \scalebox{0.8}{\scorebw{5}{5}} \\ \hline
\end{tabular}%
}
  \label{tab:prompt_and_feedback}%
  \vspace{-5pt}
\end{table}

Non-determinism is a fundamental challenge underlying the instability of LLM-generated outputs\cite{chen2021LLMrandom}\cite{ouyang2025empirical}. 
Structured prompting\cite{liu2023VerilogEval}\cite{lu2024RTLLM} improves the likelihood of syntactically valid outputs by reducing ambiguity, while feedback-based supervision\cite{Thakur2024AutoChip} enforces correctness through error detection and correction.
However, during the reasoning process of LLMs, there is often a lack of effective reasoning path planning—while goals may be specified, the intermediate steps are typically unguided or loosely inferred, leading to inconsistencies in reasoning. Chain-of-Thought (CoT) prompting, which encourages LLMs to reason step by step through intermediate reasoning traces, has been employed to improve reasoning in non-HDL code generation \cite{li2025structured}. However, its application in hardware design remains largely unexplored.
A comparative summary of these methods in terms of prompting strategy, feedback mechanism, LLM non-determinism, and output quality is provided in Table \ref{tab:prompt_and_feedback}. 
By combining these approaches, their strengths are leveraged to enhance the overall quality and stability of the generated code.

\begin{figure}[h]
    \centering
    \vspace{-10pt}
    \includegraphics[width=0.95\linewidth]{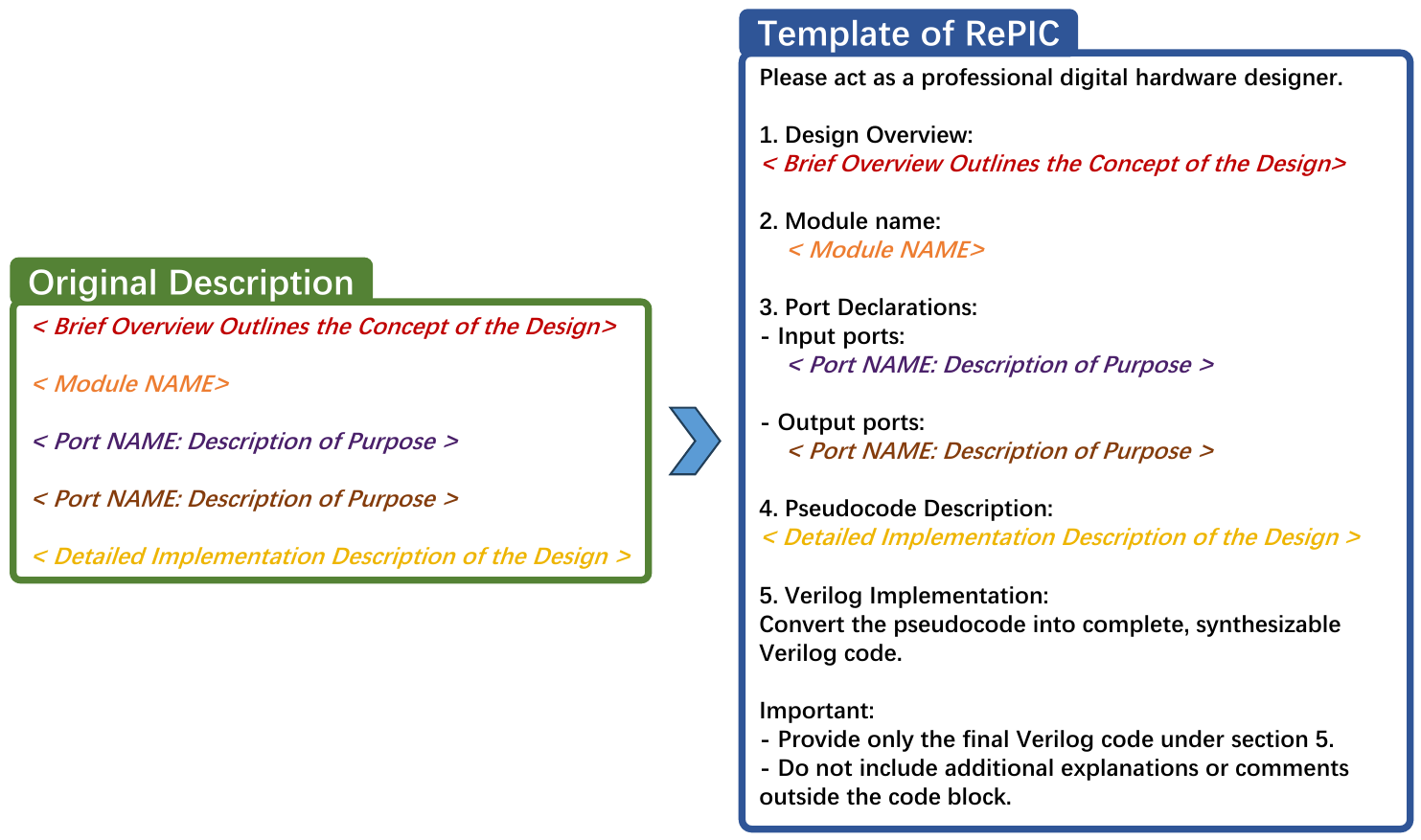}
    \caption{The original description (left) and the template of RePIC (right).}
    \label{fig:input_prompt}
    \vspace{-10pt}
\end{figure}

\begin{figure}[h]
    \centering
    \vspace{-10pt}    \includegraphics[width=0.75\linewidth]{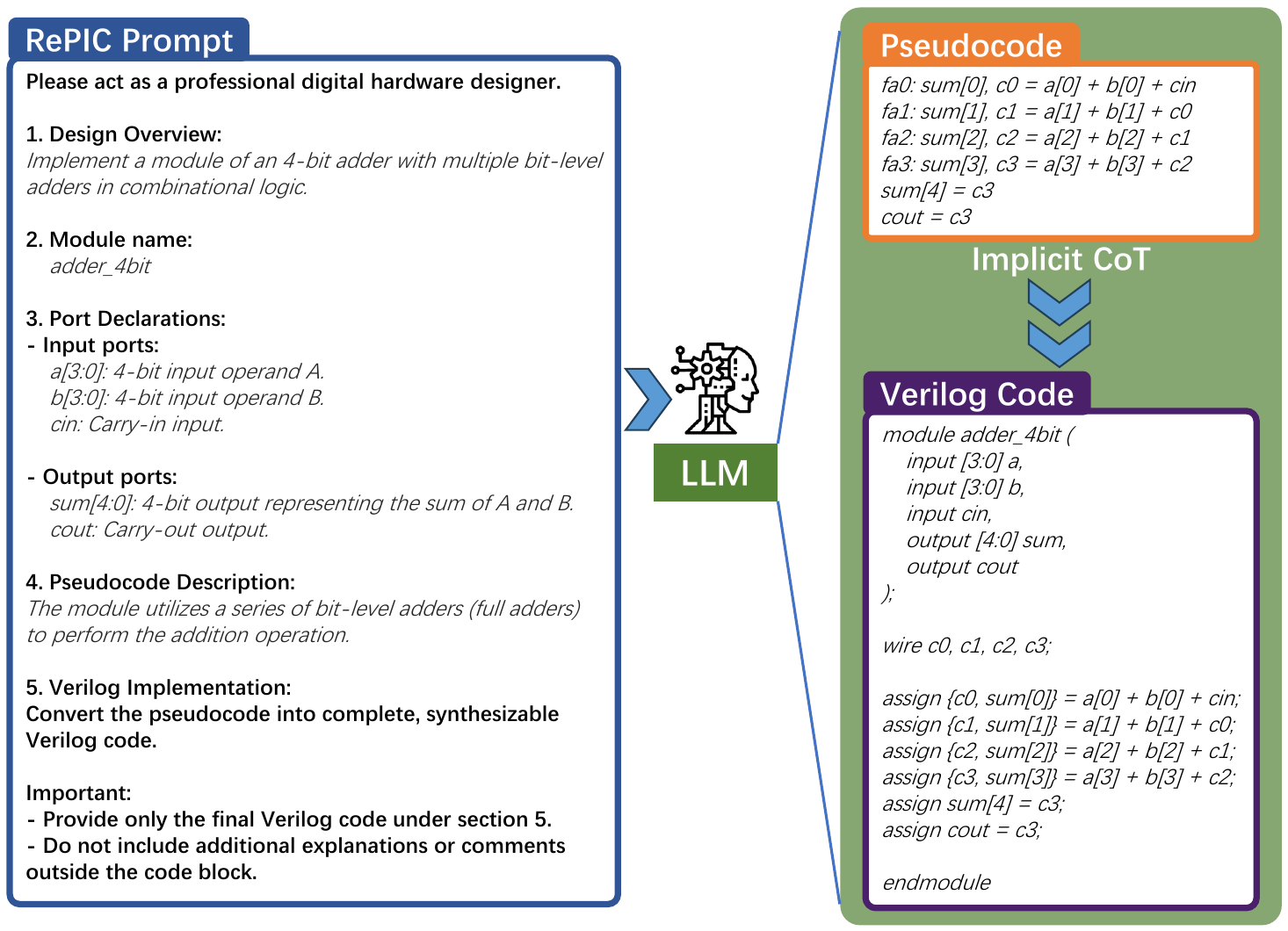}
    \caption{RePIC pipeline: from RePIC prompt to Implicit CoT-guided Verilog code generation.}
    \label{fig:Cot_example}
    \vspace{-10pt}
\end{figure}

We propose a new prompting methodology inspired by existing techniques. The first component of our methodology is 
{\bf Regulated Prompting with Implicit CoT (RePIC)}, a structured prompting strategy that introduces pseudocode as an intermediate step to guide implicit CoT reasoning. Examples of the original description and the  RePIC template are provided in Figure~\ref{fig:input_prompt}. Since Lorecast is to provide circuit designs with quick estimates of performance and power, instead of helping a layman to design circuits, its users normally have sufficient experience to supply the essential design information in natural language. This information is automatically regularized by our system into a structured prompt, as illustrated in Figure \ref{fig:input_prompt}. The regulated prompt then guides the LLM to implicitly generate pseudocode from the design description, which is subsequently converted into corresponding Verilog code, as shown in Figure \ref{fig:Cot_example}.

\begin{figure}[H]
\vspace{-6pt}
    \centering
\includegraphics[width=0.5\linewidth]{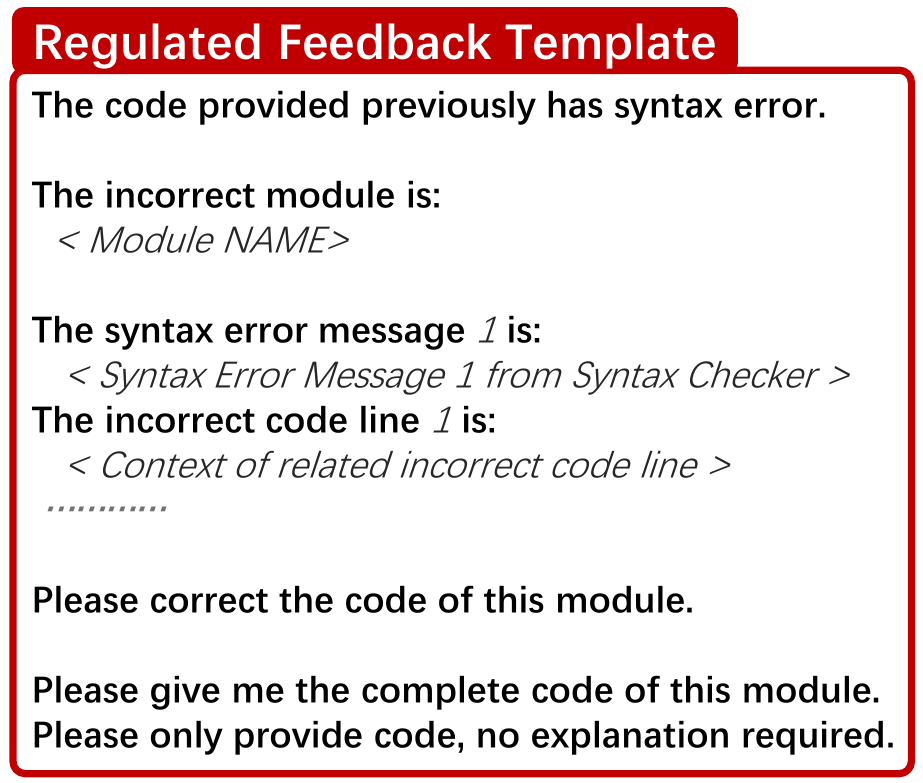}
\vspace{-10pt}
    \caption{An example of a template for regulated feedback prompting.}
\label{fig:RegulatedFeedbackPrompt}
\end{figure}

The second component is {\bf Iterative Prompting with Regulated Error Feedback (I-PREF)}. I-PREF is performed when there are syntax errors in the Verilog code generated by the LLM. Various tools are available for checking Verilog code syntax correctness, including Icarus Verilog \cite{Williams2024Icarus}, Synopsys VCS \cite{Synopsys_VCS}, Xcelium \cite{Cadence2023Xcelium}, PyVerilog \cite{Yamazak2015Pyverilog}, and Verilator \cite{Snyder2004Verilator}. 
In this work, we adopt Icarus Verilog for syntax checking. In I-PREF, the error messages along with the Verilog code with syntax errors are sent back to the LLM as a new prompt for generating updated Verilog code. This process is repeated until there is no syntax error or the maximum limit $N$ is reached. Usually, $N$ is set to be 10 as improvement can rarely be obtained after 10 iterations. The original idea of taking error feedback for iterative prompting was introduced in \cite{Thakur2024AutoChip}. However, its feedback prompts are not regulated. By contrast, we propose regulated feedback prompting and an example of such a regulation template is provided in Figure~\ref{fig:RegulatedFeedbackPrompt}.

\subsection{Phase II: Performance and Power Forecasting from Verilog Code}
\label{sec:prediction}



\begin{figure}[h]
    \centering
    \includegraphics[width=0.5\linewidth]{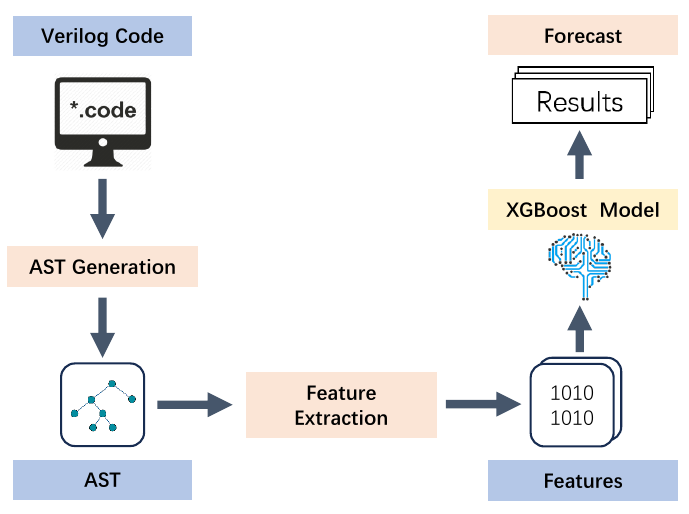}
    \caption{Performance and power forecasting from Verilog code.}
    \label{fig:prediction_flow}
    \vspace{-10pt}
\end{figure}

In Phase II, the Verilog code generated from Phase I is utilized for performance and power forecasting. In general, the code should have no syntax errors. On the other hand, our method tolerates limited functional errors, i.e., even if the code cannot be synthesized to correct circuits, it can still be applied to provide reasonable performance/power estimates. Several previous works have attempted to make performance and power predictions based on Verilog code \cite{Sengupta2022Prediction}\cite{Xu2022systhesisPrediction}\cite{Fang2023MasterRTL}\cite{Lopera2023GNNPrediction}, and we adopt the approach from \cite{Sengupta2022Prediction} with one modification. The models of \cite{Sengupta2022Prediction} are trained on post-placement analysis data while the models used by Lorecast are trained on post-routing analysis data. Thus, Lorecast is expected to provide a more accurate forecast in capturing the layout impact.
As shown in Figure~\ref{fig:prediction_flow}, the input Verilog code is first parsed to obtain an Abstract Syntax Tree (AST) using an off-the-shelf software tool\cite{Yamazak2015Pyverilog}. Next, features are extracted from the obtained AST. Then, an XGBoost model is applied to obtain the performance and power forecast for the corresponding Verilog code with the AST features and EDA tool parameters as input.


\subsection{Functional Correctness versus Structural Similarity}
\label{sec:structural_similarity}

\begin{figure}[h] 
\centering 
\includegraphics[width=0.9\linewidth]{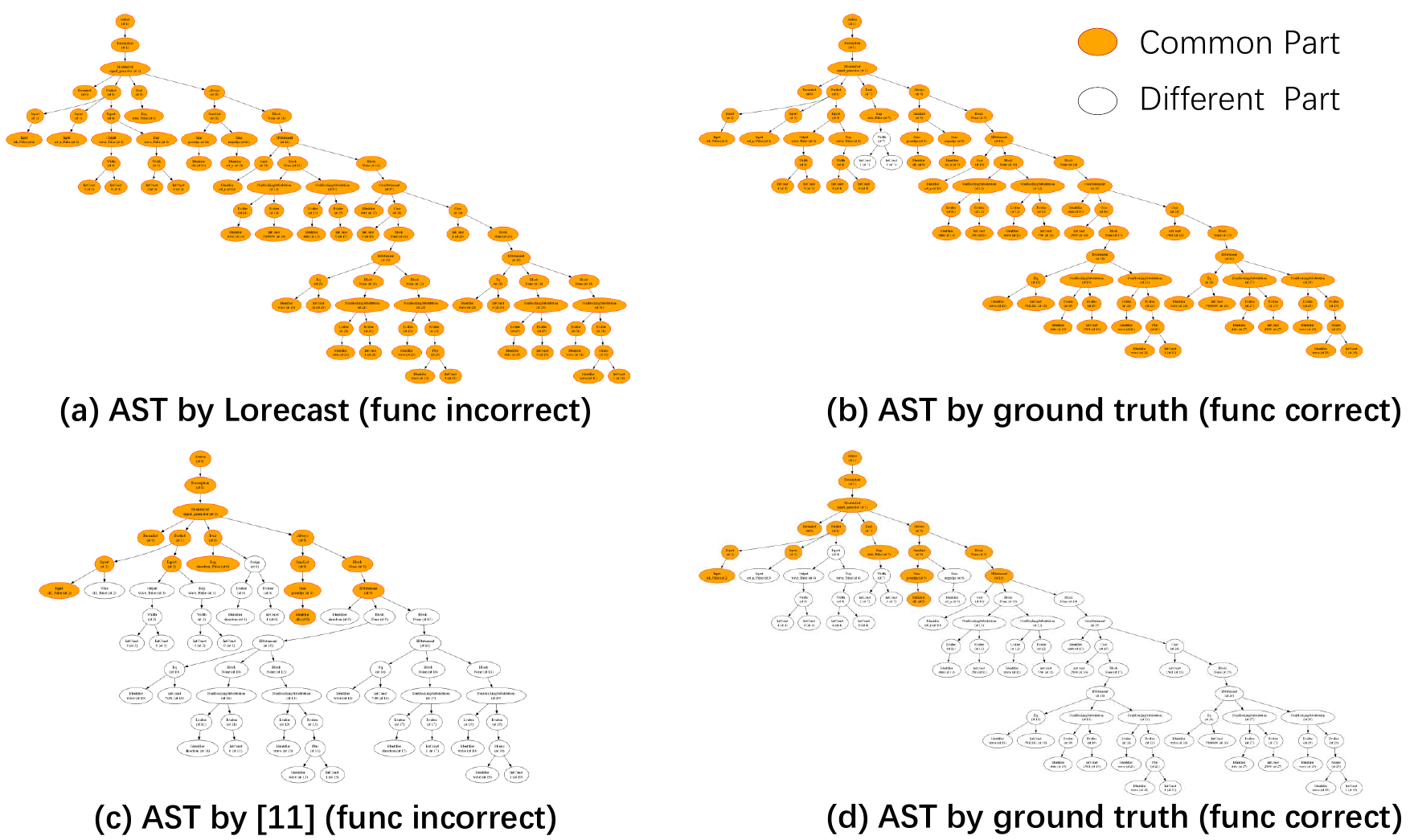}
\caption{ASTs of functionally correct and incorrect Verilog code for the same testcase. (d) same as (b). Orange color highlights common subtrees between (a)-(b) and (c)-(d).}
\label{fig:AST_common}
\vspace{-10pt}
\end{figure}

Lorecast requires the LLM-generated Verilog code to be syntactically correct but not necessarily functionally correct. Why, then, can Lorecast still be effective despite functional inaccuracies? A key reason lies in the AST structure of the generated code. The Verilog code produced by Lorecast not only achieves a high rate of syntax correctness but also exhibits an AST structure that closely resembles that of functionally correct Verilog code. This is illustrated in Figure~\ref{fig:AST_common}: the AST of Lorecast-generated code in (a) is highly similar to the AST of the functionally correct Verilog code in (b). In contrast, the AST produced by another code generator~\cite{liu2023VerilogEval} in (c) differs significantly from that of the correct version shown in (d).

  

  


\section{Experimental Results}
\label{sec:Experiment}

\subsection{Experiment Setup}

\noindent{\bf Testcases.}
In addition to the dataset of RTLLM~\cite{lu2024RTLLM}, we have prepared new designs, many of which are larger than
those in \cite{lu2024RTLLM}, for training the XGBoost model and testing the techniques. The testcases are distinct from the designs in the training dataset. Seven of the testcases are from \cite{lu2024RTLLM}, one is from \cite{haoxing2025VerilogEvalv2}, and the other seven are newly produced by us. Please note that most testcases from \cite{haoxing2025VerilogEvalv2} are very small, with a median of only 10 cells, and repetitions with small variances. 
The statistics of our testcases are shown in 
Table~\ref{tab:testset} in comparison with testcases of previous works. As shown, our testcases are significantly larger than the previous ones. Moreover, our cases cover a wide variety of designs, including datapath designs, such as multipliers and FFT units, and control logic intensive circuits, such as signal generator and Huffman decoder.


\begin{table}[h]
\centering
\vspace{-12pt}
\caption{Cell count statistics of testcases used by
Lorecast in comparison with testcases of related works.
}
\resizebox{0.75\linewidth}{!}{%
\begin{tabular}{|c||c||ccc|}
\hline
\multirow{2}{*}{Work} & \multirow{2}{*}{\begin{tabular}[c]{@{}c@{}}Num of\\ designs\end{tabular}} & \multicolumn{3}{c|}{Number of cells} \\ \cline{3-5} 
 &  & Median & Mean & Max \\ \hline
 \hline
Chip-Chat~\cite{Blocklove2023Chip-Chat} & 8 & 37 & 44 & 110 \\
Thakur, et al.\cite{Thakur2023Benchmarking} & 17 & 9.5 & 45 & 335 \\
RTLLM~\cite{lu2024RTLLM} & 30 & 121 & 408 & 2435 \\ 
\hline
\hline
Lorecast testcases & 15 & 890 & 1806 & 12031 \\ \hline
\end{tabular}%
}
  \label{tab:testset}%
  \vspace{-10pt}
\end{table}

\vspace*{1mm}
\noindent{\bf Techniques evaluated.}
Several LLMs are evaluated, including GPT3.5, GPT4, GPT4o~\cite{OpenAI2023GPT-4}, Llama3, Llama3.1~\cite{meta2024Llama}, Gemini1.5, Gemini1.5Pro~\cite{Google2024Gemini}, and DeepSeek V3. Lorecast is examined with GPT4, GPT4o,  Gemini1.5Pro, and DeepSeek V3, the four best-performing LLMs for Verilog code generation.
In addition, the RePIC technique and I-PREF technique (Section~\ref{sec:code_generation}) are also assessed. 

\vspace*{1mm}
\noindent{\bf Metrics for evaluating Verilog code generation.}
In the experiments, we evaluate syntax correctness,
functional correctness of the generated Verilog code, and the accuracy of Lorecast forecasting.
Syntax checking for LLM-generated Verilog code is performed using Icarus~\cite{Williams2024Icarus}. Functional correctness is assessed through RTL simulation using Icarus. In previous works~\cite{liu2023VerilogEval,lu2024RTLLM}, syntax/functional correctness is evaluated by $pass@k$, which means the probability of any syntax/functional correct generation in $k$ attempts for a design. Such a metric generally means multiple attempts are needed to produce syntax/functionally correct code. Since the goal of Lorecast is to obtain a quick estimation and we try to avoid multiple attempts. Hence, we adopt a simpler and significantly stricter metric. For each design,
we report {\bf correct} if all attempts, whether single or multiple, lead to syntax/functional correctness, i.e., the ``correct" here is a binary indicator instead of probability. 
We also report the {\bf correct rate}, which is the ratio of the number of designs where the generated codes are correct versus the total number of designs in the testcases.

\vspace*{1mm}
\noindent
{\bf Forecast accuracy metrics.}
The accuracy of TNS and power forecast is evaluated by 
{\bf Absolute Percentage Mean Error (APME)}. All TNS values are reported as their absolute values.
Let $\bar{y}$ be the average forecast result among all designs, and $\bar{y}^*$ be the ground truth average among all designs. Then, APME is defined by
\begin{equation}
\mathcal{E} = \frac{|\bar{y}-\bar{y}^*|}{\bar{y}^*}\times 100\% \label{eq:APME}
\end{equation}
The reason that we could not use Mean Absolute Percentage Error (MAPE) is that some ground truth TNS values are 0. 
We also report the \textbf{Normalized Root Mean Square Error (NRMSE)}. Due to large variations in data magnitude, RMSE can be hard to interpret directly. To address this, we normalize it using the mean of the ground truth values. NRMSE is defined as
\begin{equation}
\mathcal{E}_{\text{NRMSE}} = \frac{\sqrt{\frac{1}{n} \sum_{i=1}^{n} (y_i - \hat{y}_i)^2}}{\bar{y}} \times 100\%
\label{eq:RNSME}
\end{equation}
In addition, the accuracy is assessed by the
$R^2$ correlation factor.
In comparison with other workflows, we observe that the others often fail to produce valid forecasts due to syntax errors in the generated designs. To quantify this, we define the Syntax Correctness Rate, $ \mathcal{\rho} _{\mathrm{syntax}}$, as the ratio of syntactically correct designs to the total number of generated designs. Furthermore, we compute Conditional Accuracy and Conditional Error for the subset of designs with correct syntax, as defined below
\begin{equation}
\mathcal{A} _{\mathrm{cond}}=\mathcal{\rho} _{\mathrm{syntax}}\cdot \left( 1-\mathcal{E} \right) \times 100\%
\label{eq:conditional_accuracy}
\end{equation}
\begin{equation}
\mathcal{E} _{cond}=1-\mathcal{A} _{\mathrm{cond}}=1-\left( \mathcal{\rho} _{\mathrm{syntax}}\cdot \left( 1-\mathcal{E} \right) \right) \times 100\%
\label{eq:conditional_error}
\end{equation}

\vspace*{1mm}
\noindent{\bf Ground truth and computing platform.}
All ground truth data are based on manually written and functionally correct Verilog code. Logic synthesis is performed on the codes by Synopsys Design Compiler with a 45nm cell library~\cite{NanGateOpenCellLibrary}. The layout, including placement and routing, is obtained using Cadence Innovus. The ground truth timing and power results are obtained through post-layout analysis. 
Logic and layout synthesis run on a Linux x86\_64 machine with AMD EPYC 7443 24-Core processors (48 cores in total), while ML model predictions are performed on a Windows 10 computer with an 11th Gen Intel(R) Core(TM) i7-11800H processor at 2.30GHz and 32GB RAM.



\subsection{Main Results}

\begin{table*}[ht]
\centering
\caption{Performance and power forecasting results.}
\resizebox{\textwidth}{!}{
\begin{tabular}{l||rr||rr||ccrr||ccrr||ccrr||ccrr}
\hline
\multicolumn{1}{c||}{\multirow{2}{*}{\textbf{Design}}} & \multicolumn{2}{c||}{\textbf{Ground Truth{+}}} & \multicolumn{2}{c||}{\textbf{Forecast from manual}} & \multicolumn{4}{c||}{\textbf{Lorecast with GPT4}} & \multicolumn{4}{c||}{\textbf{Lorecast with Gemini 1.5 Pro}} & \multicolumn{4}{c||}{\textbf{Lorecast with GPT4o}} & \multicolumn{4}{c}{\textbf{Lorecast with DeepSeek V3}} \\
\multicolumn{1}{c||}{} & \multicolumn{1}{c|}{Power{*}} & \multicolumn{1}{c||}{TNS{*}} & \multicolumn{1}{c|}{Power} & \multicolumn{1}{c||}{TNS} & \multicolumn{1}{c|}{Syntax} & \multicolumn{1}{c|}{Func} & \multicolumn{1}{c|}{Power} & \multicolumn{1}{c||}{TNS} & \multicolumn{1}{c|}{Syntax} & \multicolumn{1}{c|}{Func} & \multicolumn{1}{c|}{Power} & \multicolumn{1}{c||}{TNS} & \multicolumn{1}{c|}{Syntax} & \multicolumn{1}{c|}{Func} & \multicolumn{1}{c|}{Power} & \multicolumn{1}{c||}{TNS} & \multicolumn{1}{c|}{Syntax} & \multicolumn{1}{c|}{Func} & \multicolumn{1}{c|}{Power} & \multicolumn{1}{c}{TNS} \\ \hline
\hline
right\_shifter & \multicolumn{1}{r|}{992} & 0.069 & \multicolumn{1}{r|}{963} & 0.0691 & \multicolumn{1}{c|}{\ding{51}} & \multicolumn{1}{c|}{\ding{51}} & \multicolumn{1}{r|}{748} & 0.0689 & \multicolumn{1}{c|}{\ding{51}} & \multicolumn{1}{c|}{\ding{51}} & \multicolumn{1}{r|}{927} & 0.0689 & \multicolumn{1}{c|}{\ding{51}} & \multicolumn{1}{c|}{\ding{51}} & \multicolumn{1}{r|}{748} & 0.0689 & \multicolumn{1}{c|}{\ding{51}} & \multicolumn{1}{c|}{\ding{51}} & \multicolumn{1}{r|}{683} & 0.0711 \\
adder\_bcd & \multicolumn{1}{r|}{8} & 0 & \multicolumn{1}{r|}{22} & 0.0200 & \multicolumn{1}{c|}{\ding{51}} & \multicolumn{1}{c|}{\ding{51}} & \multicolumn{1}{r|}{24} & 0 & \multicolumn{1}{c|}{\ding{51}} & \multicolumn{1}{c|}{\ding{51}} & \multicolumn{1}{r|}{23} & 0 & \multicolumn{1}{c|}{\ding{51}} & \multicolumn{1}{c|}{\ding{51}} & \multicolumn{1}{r|}{25} & 0 & \multicolumn{1}{c|}{\ding{51}} & \multicolumn{1}{c|}{\ding{51}} & \multicolumn{1}{r|}{26} & 0.0014 \\
signal\_generator & \multicolumn{1}{r|}{1,508} & 0.225 & \multicolumn{1}{r|}{1,347} & 0.2249 & \multicolumn{1}{c|}{\ding{51}} & \multicolumn{1}{c|}{\ding{51}} & \multicolumn{1}{r|}{1,547} & 0.2135 & \multicolumn{1}{c|}{\ding{51}} & \multicolumn{1}{c|}{\ding{55}} & \multicolumn{1}{r|}{1,547} & 0.2135 & \multicolumn{1}{c|}{\ding{51}} & \multicolumn{1}{c|}{\ding{55}} & \multicolumn{1}{r|}{1,547} & 0.2184 & \multicolumn{1}{c|}{\ding{51}} & \multicolumn{1}{c|}{\ding{55}} & \multicolumn{1}{r|}{1,547} & 0.2184 \\
multiply & \multicolumn{1}{r|}{35} & 0 & \multicolumn{1}{r|}{26} & 0.0002 & \multicolumn{1}{c|}{\ding{51}} & \multicolumn{1}{c|}{\ding{51}} & \multicolumn{1}{r|}{37} & 0.0001 & \multicolumn{1}{c|}{\ding{51}} & \multicolumn{1}{c|}{\ding{51}} & \multicolumn{1}{r|}{69} & 0.0015 & \multicolumn{1}{c|}{\ding{51}} & \multicolumn{1}{c|}{\ding{51}} & \multicolumn{1}{r|}{34} & 0.0008 & \multicolumn{1}{c|}{\ding{51}} & \multicolumn{1}{c|}{\ding{51}} & \multicolumn{1}{r|}{27} & 0.0027 \\
accu & \multicolumn{1}{r|}{4,042} & 0.306 & \multicolumn{1}{r|}{3,774} & 0.3043 & \multicolumn{1}{c|}{\ding{51}} & \multicolumn{1}{c|}{\ding{51}} & \multicolumn{1}{r|}{3,996} & 0.3062 & \multicolumn{1}{c|}{\ding{51}} & \multicolumn{1}{c|}{\ding{51}} & \multicolumn{1}{r|}{4,316} & 0.3049 & \multicolumn{1}{c|}{\ding{51}} & \multicolumn{1}{c|}{\ding{51}} & \multicolumn{1}{r|}{3,996} & 0.3318 & \multicolumn{1}{c|}{\ding{51}} & \multicolumn{1}{c|}{\ding{51}} & \multicolumn{1}{r|}{3,996} & 0.3318 \\
LIFObuffer & \multicolumn{1}{r|}{29} & 0 & \multicolumn{1}{r|}{39} & 0.0010 & \multicolumn{1}{c|}{\ding{51}} & \multicolumn{1}{c|}{\ding{51}} & \multicolumn{1}{r|}{33} & 0.0001 & \multicolumn{1}{c|}{\ding{51}} & \multicolumn{1}{c|}{\ding{51}} & \multicolumn{1}{r|}{44} & 0.0006 & \multicolumn{1}{c|}{\ding{51}} & \multicolumn{1}{c|}{\ding{51}} & \multicolumn{1}{r|}{48} & 0.0027 & \multicolumn{1}{c|}{\ding{51}} & \multicolumn{1}{c|}{\ding{51}} & \multicolumn{1}{r|}{20} & 0.0009 \\
asyn\_fifo & \multicolumn{1}{r|}{100} & 0 & \multicolumn{1}{r|}{106} & 0.0010 & \multicolumn{1}{c|}{\ding{51}} & \multicolumn{1}{c|}{\ding{55}} & \multicolumn{1}{r|}{160} & 0.0003 & \multicolumn{1}{c|}{\ding{51}} & \multicolumn{1}{c|}{\ding{55}} & \multicolumn{1}{r|}{210} & 0 & \multicolumn{1}{c|}{\ding{51}} & \multicolumn{1}{c|}{\ding{55}} & \multicolumn{1}{r|}{180} & 0.0003 & \multicolumn{1}{c|}{\ding{51}} & \multicolumn{1}{c|}{\ding{55}} & \multicolumn{1}{r|}{148} & 0.0003 \\
sobel\_filter & \multicolumn{1}{r|}{49,475} & 0.995 & \multicolumn{1}{r|}{49,226} & 0.9842 & \multicolumn{1}{c|}{\ding{51}} & \multicolumn{1}{c|}{\ding{55}} & \multicolumn{1}{r|}{47,813} & 0.9541 & \multicolumn{1}{c|}{\ding{51}} & \multicolumn{1}{c|}{\ding{55}} & \multicolumn{1}{r|}{51,574} & 1.0499 & \multicolumn{1}{c|}{\ding{51}} & \multicolumn{1}{c|}{\ding{55}} & \multicolumn{1}{r|}{47,088} & 0.9594 & \multicolumn{1}{c|}{\ding{51}} & \multicolumn{1}{c|}{\ding{55}} & \multicolumn{1}{r|}{50,885} & 1.0505 \\
matmul22 & \multicolumn{1}{r|}{307} & 0 & \multicolumn{1}{r|}{312} & 0.0002 & \multicolumn{1}{c|}{\ding{51}} & \multicolumn{1}{c|}{\ding{51}} & \multicolumn{1}{r|}{326} & 0 & \multicolumn{1}{c|}{\ding{51}} & \multicolumn{1}{c|}{\ding{51}} & \multicolumn{1}{r|}{376} & 0 & \multicolumn{1}{c|}{\ding{51}} & \multicolumn{1}{c|}{\ding{51}} & \multicolumn{1}{r|}{326} & 0 & \multicolumn{1}{c|}{\ding{51}} & \multicolumn{1}{c|}{\ding{51}} & \multicolumn{1}{r|}{371} & 0.0004 \\
mux256to1v & \multicolumn{1}{r|}{79} & 0 & \multicolumn{1}{r|}{36} & 0.0001 & \multicolumn{1}{c|}{\ding{51}} & \multicolumn{1}{c|}{\ding{51}} & \multicolumn{1}{r|}{45} & 0.0002 & \multicolumn{1}{c|}{\ding{51}} & \multicolumn{1}{c|}{\ding{51}} & \multicolumn{1}{r|}{32} & 0.0002 & \multicolumn{1}{c|}{\ding{51}} & \multicolumn{1}{c|}{\ding{51}} & \multicolumn{1}{r|}{32} & 0.0002 & \multicolumn{1}{c|}{\ding{51}} & \multicolumn{1}{c|}{\ding{51}} & \multicolumn{1}{r|}{43} & 0.0002 \\
pe\_32bit & \multicolumn{1}{r|}{84,818} & 1.139 & \multicolumn{1}{r|}{86,900} & 1.1384 & \multicolumn{1}{c|}{\ding{51}} & \multicolumn{1}{c|}{\ding{51}} & \multicolumn{1}{r|}{86,442} & 1.1432 & \multicolumn{1}{c|}{\ding{51}} & \multicolumn{1}{c|}{\ding{51}} & \multicolumn{1}{r|}{86,442} & 1.1432 & \multicolumn{1}{c|}{\ding{51}} & \multicolumn{1}{c|}{\ding{51}} & \multicolumn{1}{r|}{86,442} & 1.1432 & \multicolumn{1}{c|}{\ding{51}} & \multicolumn{1}{c|}{\ding{51}} & \multicolumn{1}{r|}{86,442} & 1.1432 \\
izigzag & \multicolumn{1}{r|}{221,131} & 0.05 & \multicolumn{1}{r|}{221,992} & 0.0496 & \multicolumn{1}{c|}{\ding{51}} & \multicolumn{1}{c|}{\ding{51}} & \multicolumn{1}{r|}{217,893} & 0.0501 & \multicolumn{1}{c|}{\ding{51}} & \multicolumn{1}{c|}{\ding{51}} & \multicolumn{1}{r|}{226,328} & 0.0495 & \multicolumn{1}{c|}{\ding{51}} & \multicolumn{1}{c|}{\ding{51}} & \multicolumn{1}{r|}{217,893} & 0.0501 & \multicolumn{1}{c|}{\ding{51}} & \multicolumn{1}{c|}{\ding{51}} & \multicolumn{1}{r|}{223,398} & 0.0497 \\
huffmandecode & \multicolumn{1}{r|}{37,461} & 0.877 & \multicolumn{1}{r|}{36,499} & 0.8363 & \multicolumn{1}{c|}{\ding{51}} & \multicolumn{1}{c|}{\ding{51}} & \multicolumn{1}{r|}{37,102} & 0.8751 & \multicolumn{1}{c|}{\ding{51}} & \multicolumn{1}{c|}{\ding{51}} & \multicolumn{1}{r|}{36,658} & 0.8855 & \multicolumn{1}{c|}{\ding{51}} & \multicolumn{1}{c|}{\ding{51}} & \multicolumn{1}{r|}{37,102} & 0.8751 & \multicolumn{1}{c|}{\ding{51}} & \multicolumn{1}{c|}{\ding{51}} & \multicolumn{1}{r|}{35,958} & 0.8644 \\
pe\_64bit & \multicolumn{1}{r|}{334,799} & 1.584 & \multicolumn{1}{r|}{333,869} & 1.5933 & \multicolumn{1}{c|}{\ding{51}} & \multicolumn{1}{c|}{\ding{51}} & \multicolumn{1}{r|}{329,173} & 1.5870 & \multicolumn{1}{c|}{\ding{51}} & \multicolumn{1}{c|}{\ding{51}} & \multicolumn{1}{r|}{329,173} & 1.5870 & \multicolumn{1}{c|}{\ding{51}} & \multicolumn{1}{c|}{\ding{51}} & \multicolumn{1}{r|}{329,173} & 1.5870 & \multicolumn{1}{c|}{\ding{51}} & \multicolumn{1}{c|}{\ding{51}} & \multicolumn{1}{r|}{329,173} & 1.5870 \\
fft\_16bit & \multicolumn{1}{r|}{466,462} & 0.9 & \multicolumn{1}{r|}{466,382} & 0.8741 & \multicolumn{1}{c|}{\ding{51}} & \multicolumn{1}{c|}{\ding{51}} & \multicolumn{1}{r|}{458,226} & 0.9019 & \multicolumn{1}{c|}{\ding{51}} & \multicolumn{1}{c|}{\ding{51}} & \multicolumn{1}{r|}{483,266} & 0.8551 & \multicolumn{1}{c|}{\ding{51}} & \multicolumn{1}{c|}{\ding{51}} & \multicolumn{1}{r|}{458,226} & 0.9019 & \multicolumn{1}{c|}{\ding{51}} & \multicolumn{1}{c|}{\ding{51}} & \multicolumn{1}{r|}{483,925} & 0.855 \\ \hline
\hline
Average & \multicolumn{1}{r|}{80,083} & 0.41 & \multicolumn{1}{r|}{80,080} & 0.4063 & \textbf{} & \multicolumn{1}{c|}{\textbf{}} & \multicolumn{1}{r|}{78,904} & 0.4067 & \textbf{} & \multicolumn{1}{c|}{\textbf{}} & \multicolumn{1}{r|}{81,399} & 0.4107 & \textbf{} & \multicolumn{1}{c|}{\textbf{}} & \multicolumn{1}{r|}{78,857} & 0.4093 & \multicolumn{1}{l}{} & \multicolumn{1}{l|}{} & \multicolumn{1}{r|}{81,109} & 0.4118 \\
APME & \multicolumn{1}{r|}{} &  & \multicolumn{1}{r|}{(1\%)} & (1\%) & \multicolumn{1}{l}{} & \multicolumn{1}{l|}{} & \multicolumn{1}{r|}{(1\%)} & (1\%) & \multicolumn{1}{l}{} & \multicolumn{1}{l|}{} & \multicolumn{1}{r|}{(2\%)} & (1\%) & \multicolumn{1}{l}{} & \multicolumn{1}{l|}{} & \multicolumn{1}{r|}{(2\%)} & (1\%) & \multicolumn{1}{l}{} & \multicolumn{1}{l|}{} & \multicolumn{1}{r|}{(1\%)} & (1\%) \\
NRSME & \multicolumn{1}{r|}{} &  & \multicolumn{1}{r|}{(1\%)} & (4\%) & \multicolumn{1}{l}{} & \multicolumn{1}{l|}{} & \multicolumn{1}{r|}{(3\%)} & (3\%) & \multicolumn{1}{l}{} & \multicolumn{1}{l|}{} & \multicolumn{1}{r|}{(6\%)} & (5\%) & \multicolumn{1}{l}{} & \multicolumn{1}{l|}{} & \multicolumn{1}{r|}{(4\%)} & (3\%) & \multicolumn{1}{l}{} & \multicolumn{1}{l|}{} & \multicolumn{1}{r|}{(6\%)} & (5\%) \\ \hline
\multicolumn{16}{l}{$^{\mathrm{+}}$Ground truth is obtained by manually written and functionally correct Verilog code, followed by logic synthesis and post-routing analysis.}\\
\multicolumn{16}{l}{$^{\mathrm{*}}$Power unit is $\mu W$ and TNS  unit is $ns$.}
\end{tabular}
}
  \label{tab:maintable}%
  \vspace{-10pt}
\end{table*}

\begin{table*}[!htb]
\centering
\caption{Time cost{*} of Verilog code writing/debug + synthesis in comparison with Lorecast.}
\resizebox{0.9\textwidth}{!}{%
\begin{tabular}{l||cccc||cccc||cccc}
\hline
\multicolumn{1}{c||}{\multirow{2}{*}{\textbf{Design}}} & \multicolumn{4}{c||}{\textbf{Manual   Verilog + synthesis}} & \multicolumn{4}{c||}{\textbf{Lorecast}} & \multicolumn{4}{c}{\textbf{Speedup}} \\ 
\multicolumn{1}{c||}{} & \multicolumn{1}{c|}{Verilog coding} & \multicolumn{1}{c|}{Logic} & \multicolumn{1}{c|}{Layout} & Total & \multicolumn{1}{c|}{Prompt writing} & \multicolumn{1}{c|}{\begin{tabular}[c]{@{}c@{}}LLM code \\ generation\end{tabular}} & \multicolumn{1}{c|}{ML prediction} & Total & \multicolumn{1}{c|}{\begin{tabular}[c]{@{}c@{}}	Manual \\ time \end{tabular}} & \multicolumn{1}{c|}{\begin{tabular}[c]{@{}c@{}} CPU\\time\end{tabular}} & \multicolumn{1}{c|}{\begin{tabular}[c]{@{}c@{}}	End-to-end\\time \end{tabular}} & \begin{tabular}[c]{@{}c@{}}Incremental change time\\(synthesis param update only)\end{tabular} \\ \hline
\hline
right\_shifter & \multicolumn{1}{c|}{1489} & 215 & \multicolumn{1}{c|}{37} & 1741 & \multicolumn{1}{c|}{270} & 5 & \multicolumn{1}{c|}{0.0017} & 275 & 5.5X & 50.4X & \multicolumn{1}{c|}{6.3X} & $148\times 10^3\,\text{X}$ \\
adder\_bcd & \multicolumn{1}{c|}{1120} & 223 & \multicolumn{1}{c|}{43} & 1386 & \multicolumn{1}{c|}{420} & 5 & \multicolumn{1}{c|}{0.0013} & 425 & 2.7X & 53.2X & \multicolumn{1}{c|}{3.3X} & $205\times 10^3\,\text{X}$ \\
signal\_generator & \multicolumn{1}{c|}{1990} & 220 & \multicolumn{1}{c|}{38} & 2248 & \multicolumn{1}{c|}{439} & 6 & \multicolumn{1}{c|}{0.0164} & 445 & 4.5X & 43.0X & \multicolumn{1}{c|}{5.1X} & $16\times 10^3\,\text{X}$ \\
accu & \multicolumn{1}{c|}{1129} & 223 & \multicolumn{1}{c|}{55} & 1407 & \multicolumn{1}{c|}{250} & 8 & \multicolumn{1}{c|}{0.023} & 258 & 4.5X & 34.8X & \multicolumn{1}{c|}{5.5X} & $12\times 10^3\,\text{X}$ \\
LIFObuffer & \multicolumn{1}{c|}{1640} & 223 & \multicolumn{1}{c|}{44} & 1907 & \multicolumn{1}{c|}{560} & 9 & \multicolumn{1}{c|}{0.0021} & 569 & 2.9X & 29.7X & \multicolumn{1}{c|}{3.4X} & $127\times 10^3\,\text{X}$ \\
multiply & \multicolumn{1}{c|}{1780} & 217 & \multicolumn{1}{c|}{40} & 2037 & \multicolumn{1}{c|}{287} & 5 & \multicolumn{1}{c|}{0.0243} & 292 & 6.2X & 51.4X & \multicolumn{1}{c|}{7.0X} & $11\times 10^3\,\text{X}$ \\
asyn\_fifo & \multicolumn{1}{c|}{5320} & 225 & \multicolumn{1}{c|}{41} & 5586 & \multicolumn{1}{c|}{1030} & 21 & \multicolumn{1}{c|}{0.0259} & 1051 & 5.2X & 12.7X & \multicolumn{1}{c|}{5.3X} & $10\times 10^3\,\text{X}$ \\
sobel\_filter & \multicolumn{1}{c|}{2820} & 235 & \multicolumn{1}{c|}{51} & 3106 & \multicolumn{1}{c|}{660} & 26 & \multicolumn{1}{c|}{0.022} & 686 & 4.3X & 11.0X & \multicolumn{1}{c|}{4.5X} & $13\times 10^3\,\text{X}$ \\
matmul22 & \multicolumn{1}{c|}{2459} & 224 & \multicolumn{1}{c|}{41} & 2724 & \multicolumn{1}{c|}{490} & 9 & \multicolumn{1}{c|}{0.0285} & 499 & 5.0X & 29.4X & \multicolumn{1}{c|}{5.5X} & $9\times 10^3\,\text{X}$ \\
mux256to1v & \multicolumn{1}{c|}{739} & 221 & \multicolumn{1}{c|}{49} & 1009 & \multicolumn{1}{c|}{320} & 3 & \multicolumn{1}{c|}{0.0014} & 323 & 2.3X & 90X & \multicolumn{1}{c|}{3.1X} & $193\times 10^3\,\text{X}$ \\
pe\_32bit & \multicolumn{1}{c|}{1650} & 231 & \multicolumn{1}{c|}{52} & 1933 & \multicolumn{1}{c|}{469} & 5 & \multicolumn{1}{c|}{0.0292} & 474 & 3.5X & 56.7X & \multicolumn{1}{c|}{4.1X} & $10\times 10^3\,\text{X}$ \\
huffmandecode & \multicolumn{1}{c|}{2200} & 270 & \multicolumn{1}{c|}{56} & 2526 & \multicolumn{1}{c|}{550} & 32 & \multicolumn{1}{c|}{0.0089} & 582 & 4X & 10.2X & \multicolumn{1}{c|}{4.3X} & $37\times 10^3\,\text{X}$ \\
izigzag & \multicolumn{1}{c|}{1819} & 265 & \multicolumn{1}{c|}{49} & 2133 & \multicolumn{1}{c|}{480} & 38 & \multicolumn{1}{c|}{0.0132} & 518 & 3.8X & 8.3X & \multicolumn{1}{c|}{4.1X} & $24\times 10^3\,\text{X}$ \\
pe\_64bit & \multicolumn{1}{c|}{1588} & 252 & \multicolumn{1}{c|}{69} & 1909 & \multicolumn{1}{c|}{512} & 5 & \multicolumn{1}{c|}{0.013} & 517 & 3.1X & 64.2X & \multicolumn{1}{c|}{3.7X} & $25\times 10^3\,\text{X}$ \\
fft 16bit & \multicolumn{1}{c|}{3600} & 343 & \multicolumn{1}{c|}{94} & 4037 & \multicolumn{1}{c|}{990} & 52 & \multicolumn{1}{c|}{0.0247} & 1060 & 3.6X & 8.4X & \multicolumn{1}{c|}{3.8X} & $17\times 10^3\,\text{X}$ \\ \hline
\hline
Average & \multicolumn{1}{c|}{2090} & 239 & \multicolumn{1}{c|}{51} & 2379 & \multicolumn{1}{c|}{515} & 15 & \multicolumn{1}{c|}{0.0157} & 530 & 4X & 19.3X & \multicolumn{1}{c|}{4.6X} & $57\times 10^3\,\text{X}$ \\ \hline
\multicolumn{10}{l}{$^{\mathrm{*}}$The time unit is $s$. }
\end{tabular}%
}
  \label{tab:speedtable}%
  \vspace{-5pt}
\end{table*}

The main results of performance and power forecasting are shown in Table~\ref{tab:maintable}.  
In addition, 
the table includes syntax and functional correctness results from different LLMs, where one attempt is made for each design. Columns 4 and 5 are the forecasting results from manually written functionally correct Verilog code. Using the RePIC and I-PREF, all four LLMs can achieve $100\%$ syntax correctness, which is required for Lorecast. None of the LLMs delivered $100\%$ functional correctness as expected. Among them, 
GPT4 with RePIC and I-PREF achieves 
$87\%$ functional correctness rate,
aligns with or improves upon the state-of-the-art methods~\cite{lu2024RTLLM}~\cite{Gao2024AutoVCoder}.
GPT4-based Lorecast also achieves the lowest APME.

\begin{figure}[bh]
    \centering
    \vspace{-7pt}
    \includegraphics[width=1\linewidth]{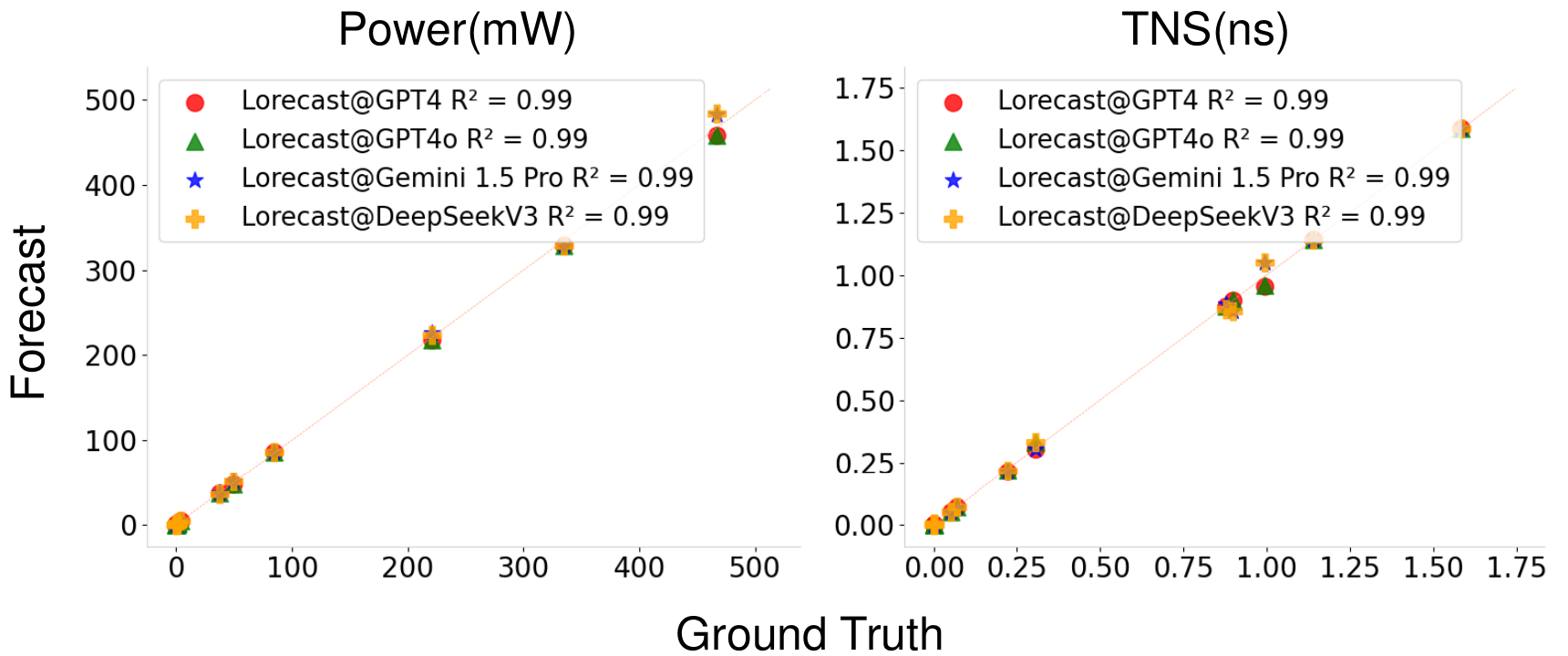}
    \vspace{-10pt}
    \caption{Power and TNS forecasts from different LLMs vs. ground truth.}
    \label{fig:scatter}
    \vspace{-5pt}
\end{figure}

Figure~\ref{fig:scatter} provides the scatter plots of forecast versus ground truth. GPT4-based Lorecast achieves $R^2$ correlations of 0.99 for power and TNS forecast, respectively. 

\begin{figure}[]
    \centering    \includegraphics[width=0.99\linewidth]{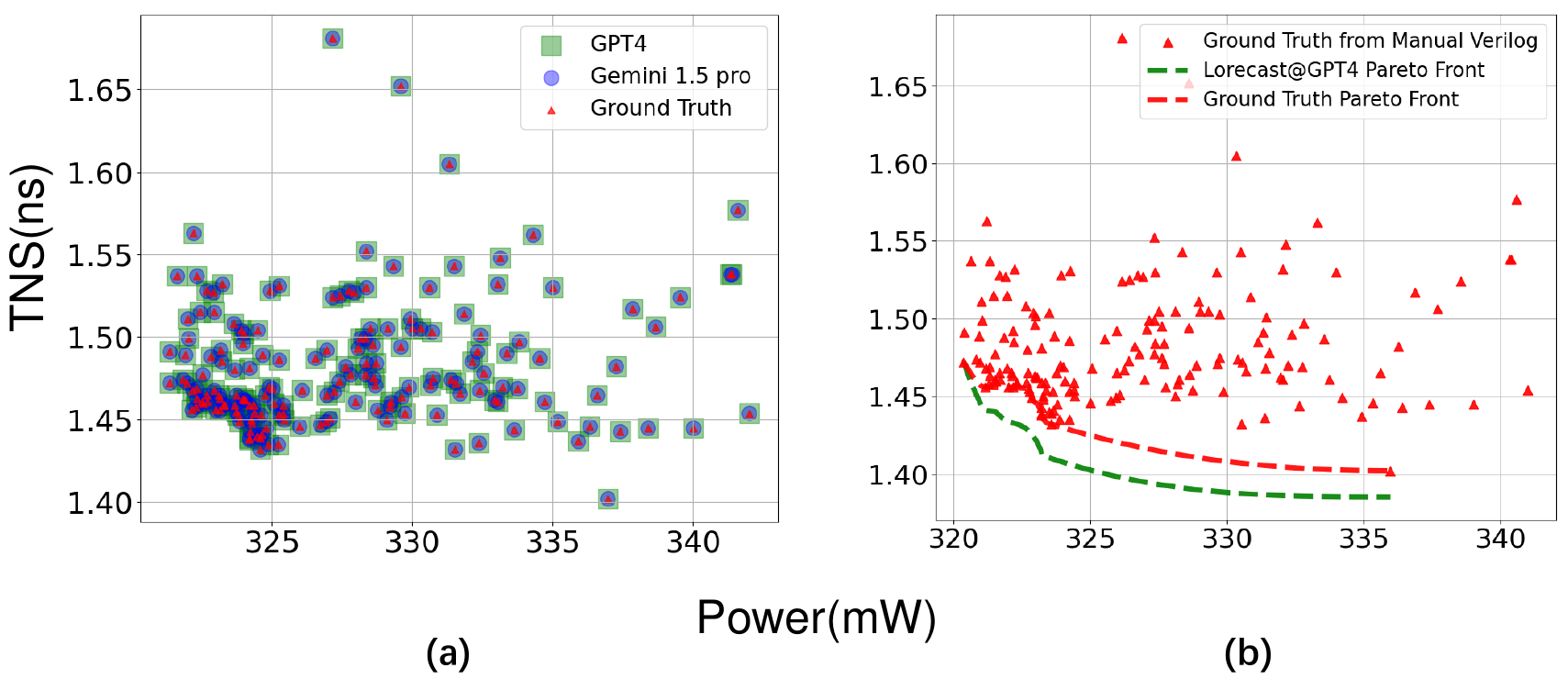}
    \vspace{-11pt}
    \caption{(a): post-layout TNS and power from Verilog code generated by LLMs and ground truth. (b): performance-power tradeoff curves predicted by Lorecast@GPT4 and the ground truth from manually written Verilog code.
}
    \label{fig:layout-results}
    \vspace{-15pt}
\end{figure}

Beyond accuracy metrics, we further examine the alignment between generated and ground truth behaviors. Figure \ref{fig:layout-results}(a) demonstrates the impact of varying logic/layout synthesis tool parameters for design “pe\_64bit”. Since the post-layout analysis of the LLM-generated code highly overlaps with the ground truth under different tool parameters, there is a consistency between the LLM-generated code and the manually written code. Figure \ref{fig:layout-results}(b) illustrates that Lorecast can predict the overall timing-power tradeoff curve, besides individual timing-power values.

\begin{figure}[] 
\vspace{-5pt}
\centering 
\includegraphics[width=0.99\linewidth]{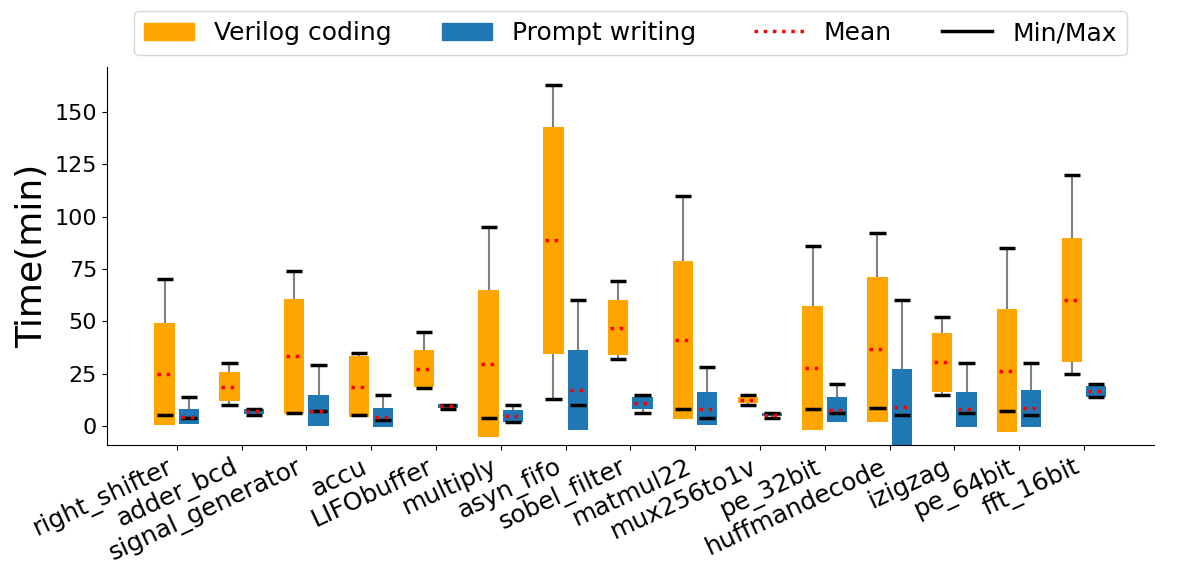}
\vspace{-15pt}
\caption{Verilog code writing and debug time in yellow bars; prompt writing and check time in blue bars. Each bar indicates $\mu\pm \sigma$ range, where $\mu$ is the mean
and $\sigma$ is the standard deviation.}
\vspace{-11pt}
\label{fig:speeddetails}
\end{figure}

\vspace{5 pt}
\noindent
{\bf Runtime comparison between Lorecast and RTL-synthesis based estimation.} 
We compared the time cost of two different flows: one is the performance/power estimation based on manually written Verilog code and logic/layout synthesis, and the other is the manually written prompts followed by Lorecast. 
The Verilog coding/debug and prompt writing/debug time are estimated by asking six participants to do both for each design. Please note that these participants generally know Verilog coding and prompting, but with different skill levels. Figure~\ref{fig:speeddetails} presents the mean, min, max, and standard deviation range of the time for all the testcase designs. Due to the limited sample size and data skew, the mean minus one standard deviation occasionally falls below the observed minimum. Overall, prompt writing/debug time is significantly shorter than Verilog code writing/debug time. It also has a smaller variance, so that time budgeting becomes easy. 


\begin{table*}[h]
\caption{Results of LLM-based reasoning without generating Verilog code.}
\centering
\vspace{-2pt}
\resizebox{0.68\linewidth}{!}{%
\begin{tabular}{l||cc|cc|cc|cc|cc|cc}
\hline
\multicolumn{1}{c||}{\multirow{2}{*}{\textbf{Design}}} & \multicolumn{2}{c|}{\textbf{GPT4}} & \multicolumn{2}{c|}{\textbf{GPT4o}} & \multicolumn{2}{c|}{\textbf{Llama3.1 405B}} & \multicolumn{2}{c|}{\textbf{Gemini 1.5 Pro}} & \multicolumn{2}{c|}{\textbf{DeepSeek R1\cite{guo2025deepseek}}} & \multicolumn{2}{c}{\textbf{\begin{tabular}[c]{@{}c@{}}Fine-tuned\\ Llama3 8B\end{tabular}}} \\
\multicolumn{1}{c||}{} & Power{*} & TNS{*} & Power & TNS & Power & TNS & Power & TNS & Power & TNS & Power & TNS \\ \hline
\hline
right\_shifter & 20 & 0.005 & 5.4 & 0.02 & 20 & 0.1 & \ding{55} & \ding{55} & 65 & 0.25 & 113 & 0 \\
adder\_bcd & 1.44 & 0 & 8.4 & 0 & 35 & 0.37 & 0.6 & 0.2 & 42 & 0.45 & 90 & 0 \\
signal\_generator & 10 & 0 & 6.8 & 0.015 & 33 & 0.02 & 5 & 0.5 & 85 & 0.35 & 103 & 0 \\
multiply & 20 & 0.025 & 8.2 & 0.018 & 76 & 0.05 & 20 & 2 & 120 & 0.35 & 105 & 4.5 \\
accu & 20 & 0.04 & 7.5 & 0.012 & 49 & 0.03 & 10 & 1 & 95 & 0.62 & 113 & 0.6 \\
LIFObuffer & 5 & 0 & 46.2 & 0 & 29 & 1.03 & 2 & 0.3 & 78 & 0.32 & 125 & 0.5 \\
asyn\_fifo & 30 & 0.005 & 9.3 & 0.02 & 128 & 0.07 & 30 & 3 & 220 & 1.2 & 120 & 6.8 \\
sobel\_filter & 30 & 2 & 93.6 & 0 & 51 & 2.51 & 10 & 0.8 & 115 & 0.58 & 310 & 8.2 \\
matmul22 & 40 & 0.035 & 10.1 & 0.025 & 262 & 0.1 & 50 & 5 & 380 & 1.8 & 240 & 3.4 \\
mux256to1v & 10 & 0 & 12.8 & 0 & 0.38 & 0.23 & 25 & 1.2 & 155 & 0.67 & 110 & 0.5 \\
pe\_32bit & 30 & 0.035 & 12.7 & 0.03 & 189 & 0.08 & 60 & 4 & 180 & 0.85 & 180 & 2.9 \\
huffmandecode & 30 & 0.025 & 14.3 & 0.035 & 336 & 0.12 & 40 & 6 & 115 & 0.75 & 240 & 4.1 \\
izigzag & 30 & 0.025 & 13.5 & 0.028 & 231 & 0.06 & 25 & 3 & 210 & 1.1 & 220 & 3.5 \\
pe\_64bit & 40 & 0.065 & 18.5 & 0.04 & 472 & 0.15 & 100 & 8 & 420 & 1.6 & 320 & 4.8 \\
fft\_16bit & 10 & 0 & 327.5 & 0 & 138 & 3.19 & 70 & 2.2 & 260 & 1.8 & 290 & 8.8 \\ \hline
\hline
Average & 22 & 0.15 & 40 & 0.016 & 137 & 0.54 & 32 & 2.7 & 169 & 0.85 & 179 & 3.2 \\
APME & (99\%) & (63\%) & (99\%) & (96\%) & (99\%) & (32\%) & (99\%) & (559\%) & (99\%) & (107\%) & (99\%) & (680\%) \\ \hline
\multicolumn{7}{l}{$^{\mathrm{*}}$Power unit is $\mu W$ and TNS  unit is $ns$. }
\end{tabular}%
}
  \label{tab:LLMreasonning}%
  \vspace{-10pt}
\end{table*}

\begin{table*}[!htb]
\caption{Impact of Lorecast prompting in comparison with other prompting techniques with the same Verilog code to power/TNS prediction technique. }
\centering
\vspace{-2pt}
\resizebox{0.8\textwidth}{!}{%
\begin{tabular}{l||rr||ccrr||ccrr||cccc}
\hline
\multicolumn{1}{c||}{\multirow{2}{*}{\textbf{Design}}} & \multicolumn{2}{c||}{\textbf{Ground Truth}} & \multicolumn{4}{c||}{\textbf{Lorecast}} & \multicolumn{4}{c||}{\textbf{\begin{tabular}[c]{@{}c@{}}Forecast with\\ RTLLM\cite{lu2024RTLLM} Prompting\end{tabular}}} & \multicolumn{4}{c}{\textbf{\begin{tabular}[c]{@{}c@{}}Forecast with\\ VerilogEval\cite{liu2023VerilogEval}\ Prompting\end{tabular}}} \\
\multicolumn{1}{c||}{} & \multicolumn{1}{c|}{Power{*}} & \multicolumn{1}{c||}{TNS{*}} & \multicolumn{1}{c|}{Syntax} & \multicolumn{1}{c|}{Func} & \multicolumn{1}{c|}{Power} & \multicolumn{1}{c||}{TNS} & \multicolumn{1}{c|}{Syntax} & \multicolumn{1}{c|}{Func} & \multicolumn{1}{c|}{Power} & \multicolumn{1}{c||}{TNS} & \multicolumn{1}{c|}{Syntax} & \multicolumn{1}{c|}{Func} & \multicolumn{1}{c|}{Power} & TNS \\ \hline
\hline
right\_shifter & \multicolumn{1}{r|}{992} & 0.069 & \multicolumn{1}{c|}{\ding{51}} & \multicolumn{1}{c|}{\ding{51}} & \multicolumn{1}{r|}{748} & 0.0689 & \multicolumn{1}{c|}{\ding{51}} & \multicolumn{1}{c|}{\ding{51}} & \multicolumn{1}{r|}{773} & 0.0690 & \multicolumn{1}{c|}{\ding{55}} & \multicolumn{1}{c|}{\ding{55}} & \multicolumn{1}{c|}{-} & - \\
adder\_bcd & \multicolumn{1}{r|}{8} & 0 & \multicolumn{1}{c|}{\ding{51}} & \multicolumn{1}{c|}{\ding{51}} & \multicolumn{1}{r|}{24} & 0 & \multicolumn{1}{c|}{\ding{51}} & \multicolumn{1}{c|}{\ding{51}} & \multicolumn{1}{r|}{37} & 0.0005 & \multicolumn{1}{c|}{\ding{55}} & \multicolumn{1}{c|}{\ding{55}} & \multicolumn{1}{c|}{-} & - \\
signal\_generator & \multicolumn{1}{r|}{1508} & 0.225 & \multicolumn{1}{c|}{\ding{51}} & \multicolumn{1}{c|}{\ding{51}} & \multicolumn{1}{r|}{1547} & 0.2135 & \multicolumn{1}{c|}{\ding{51}} & \multicolumn{1}{c|}{\ding{55}} & \multicolumn{1}{r|}{1346} & 0.2130 & \multicolumn{1}{c|}{\ding{51}} & \multicolumn{1}{c|}{\ding{55}} & \multicolumn{1}{r|}{4774} & \multicolumn{1}{r}{0.0433} \\
multiply & \multicolumn{1}{r|}{35} & 0 & \multicolumn{1}{c|}{\ding{51}} & \multicolumn{1}{c|}{\ding{51}} & \multicolumn{1}{r|}{37} & 0.0001 & \multicolumn{1}{c|}{\ding{51}} & \multicolumn{1}{c|}{\ding{51}} & \multicolumn{1}{r|}{48} & 0.0002 & \multicolumn{1}{c|}{\ding{55}} & \multicolumn{1}{c|}{\ding{55}} & \multicolumn{1}{c|}{-} & - \\
accu & \multicolumn{1}{r|}{4042} & 0.306 & \multicolumn{1}{c|}{\ding{51}} & \multicolumn{1}{c|}{\ding{51}} & \multicolumn{1}{r|}{3996} & 0.3062 & \multicolumn{1}{c|}{\ding{51}} & \multicolumn{1}{c|}{\ding{51}} & \multicolumn{1}{r|}{3717} & 0.2822 & \multicolumn{1}{c|}{\ding{55}} & \multicolumn{1}{c|}{\ding{55}} & \multicolumn{1}{c|}{-} & - \\
LIFObuffer & \multicolumn{1}{r|}{29} & 0 & \multicolumn{1}{c|}{\ding{51}} & \multicolumn{1}{c|}{\ding{51}} & \multicolumn{1}{r|}{33} & 0.0001 & \multicolumn{1}{c|}{\ding{55}} & \multicolumn{1}{c|}{\ding{55}} & \multicolumn{1}{c|}{-} & \multicolumn{1}{c||}{-} & \multicolumn{1}{c|}{\ding{55}} & \multicolumn{1}{c|}{\ding{55}} & \multicolumn{1}{c|}{-} & - \\
asyn\_fifo & \multicolumn{1}{r|}{100} & 0 & \multicolumn{1}{c|}{\ding{51}} & \multicolumn{1}{c|}{\ding{55}} & \multicolumn{1}{r|}{160} & 0.0003 & \multicolumn{1}{c|}{\ding{55}} & \multicolumn{1}{c|}{\ding{55}} & \multicolumn{1}{c|}{-} & \multicolumn{1}{c||}{-} & \multicolumn{1}{c|}{\ding{55}} & \multicolumn{1}{c|}{\ding{55}} & \multicolumn{1}{c|}{-} & - \\
sobel\_filter & \multicolumn{1}{r|}{49475} & 0.995 & \multicolumn{1}{c|}{\ding{51}} & \multicolumn{1}{c|}{\ding{55}} & \multicolumn{1}{r|}{47813} & 0.9541 & \multicolumn{1}{c|}{\ding{55}} & \multicolumn{1}{c|}{\ding{55}} & \multicolumn{1}{c|}{-} & \multicolumn{1}{c||}{-} & \multicolumn{1}{c|}{\ding{55}} & \multicolumn{1}{c|}{\ding{55}} & \multicolumn{1}{c|}{-} & - \\
matmul22 & \multicolumn{1}{r|}{307} & 0 & \multicolumn{1}{c|}{\ding{51}} & \multicolumn{1}{c|}{\ding{51}} & \multicolumn{1}{r|}{326} & 0 & \multicolumn{1}{c|}{\ding{55}} & \multicolumn{1}{c|}{\ding{55}} & \multicolumn{1}{c|}{-} & \multicolumn{1}{c||}{-} & \multicolumn{1}{c|}{\ding{55}} & \multicolumn{1}{c|}{\ding{55}} & \multicolumn{1}{c|}{-} & - \\
mux256to1v & \multicolumn{1}{r|}{79} & 0 & \multicolumn{1}{c|}{\ding{51}} & \multicolumn{1}{c|}{\ding{51}} & \multicolumn{1}{r|}{45} & 0.0002 & \multicolumn{1}{c|}{\ding{51}} & \multicolumn{1}{c|}{\ding{51}} & \multicolumn{1}{r|}{45} & 0.0001 & \multicolumn{1}{c|}{\ding{51}} & \multicolumn{1}{c|}{\ding{51}} & \multicolumn{1}{r|}{36} & \multicolumn{1}{r}{0.0001} \\
pe\_32bit & \multicolumn{1}{r|}{84818} & 1.139 & \multicolumn{1}{c|}{\ding{51}} & \multicolumn{1}{c|}{\ding{51}} & \multicolumn{1}{r|}{86442} & 1.1432 & \multicolumn{1}{c|}{\ding{51}} & \multicolumn{1}{c|}{\ding{51}} & \multicolumn{1}{r|}{87456} & 1.1463 & \multicolumn{1}{c|}{\ding{51}} & \multicolumn{1}{c|}{\ding{55}} & \multicolumn{1}{r|}{2003} & \multicolumn{1}{r}{0.0421} \\
izigzag & \multicolumn{1}{r|}{221131} & 0.05 & \multicolumn{1}{c|}{\ding{51}} & \multicolumn{1}{c|}{\ding{51}} & \multicolumn{1}{r|}{217893} & 0.0501 & \multicolumn{1}{c|}{\ding{55}} & \multicolumn{1}{c|}{\ding{55}} & \multicolumn{1}{c|}{-} & \multicolumn{1}{c||}{-} & \multicolumn{1}{c|}{\ding{55}} & \multicolumn{1}{c|}{\ding{55}} & \multicolumn{1}{c|}{-} & - \\
huffmandecode & \multicolumn{1}{r|}{37461} & 0.877 & \multicolumn{1}{c|}{\ding{51}} & \multicolumn{1}{c|}{\ding{51}} & \multicolumn{1}{r|}{37102} & 0.8751 & \multicolumn{1}{c|}{\ding{55}} & \multicolumn{1}{c|}{\ding{55}} & \multicolumn{1}{c|}{-} & \multicolumn{1}{c||}{-} & \multicolumn{1}{c|}{\ding{55}} & \multicolumn{1}{c|}{\ding{55}} & \multicolumn{1}{c|}{-} & - \\
pe\_64bit & \multicolumn{1}{r|}{334799} & 1.584 & \multicolumn{1}{c|}{\ding{51}} & \multicolumn{1}{c|}{\ding{51}} & \multicolumn{1}{r|}{329173} & 1.5870 & \multicolumn{1}{c|}{\ding{51}} & \multicolumn{1}{c|}{\ding{51}} & \multicolumn{1}{r|}{332914} & 1.6127 & \multicolumn{1}{c|}{\ding{55}} & \multicolumn{1}{c|}{\ding{55}} & \multicolumn{1}{c|}{-} & - \\
fft\_16bit & \multicolumn{1}{r|}{466462} & 0.9 & \multicolumn{1}{c|}{\ding{51}} & \multicolumn{1}{c|}{\ding{51}} & \multicolumn{1}{r|}{458226} & 0.9019 & \multicolumn{1}{c|}{\ding{55}} & \multicolumn{1}{c|}{\ding{55}} & \multicolumn{1}{c|}{-} & \multicolumn{1}{c||}{-} & \multicolumn{1}{c|}{\ding{55}} & \multicolumn{1}{c|}{\ding{55}} & \multicolumn{1}{c|}{-} & - \\ \hline
\hline
Conditional Error & \multicolumn{1}{l}{} & \multicolumn{1}{l|}{} & \multicolumn{1}{l}{} & \multicolumn{1}{l|}{} & \multicolumn{1}{r|}{(1\%)} & (1\%) & \multicolumn{1}{l}{} & \multicolumn{1}{l|}{} & \multicolumn{1}{r|}{(48\%)} & (48\%) & \multicolumn{1}{l}{} & \multicolumn{1}{l|}{} & \multicolumn{1}{r|}{(98\%)} & \multicolumn{1}{r}{(99\%)} \\ \hline
\multicolumn{9}{l}{- Forecast can not be performed due to syntax errors in Verilog code generated by LLMs.}\\
\multicolumn{6}{l}{$^{\mathrm{*}}$Power unit is $\mu W$ and TNS  unit is $ns$.}
\end{tabular}%
}
\label{tab:generate_comparison}%
\vspace{-15pt}
\end{table*}

Other components of the time cost are summarized in Table~\ref{tab:speedtable}, where the coding/debugging time results are the mean values. 
The ML prediction time is obtained using the model~\cite{Sengupta2022Prediction}. The time is dominated by Verilog code writing/debugging and prompt writing/debugging.
Overall, Lorecast can achieve $4.6\times$ speedup. 
Manual (Verilog coding/debugging or prompt writing/debugging) time has been reduced by $4\times$. For CPU time, LLM-based code generation and ML prediction  is $19.3\times$ faster than traditional logic and layout synthesis. For incremental design changes, where only synthesis parameters are modified, Lorecast re-inference achieves a speedup exceeding 50,000× compared to re-running traditional logic and layout synthesis.
Furthermore, Lorecast reduces the requirement for HDL coding skills for the performance/power estimation.

\noindent
{\bf Direct LLM-based forecasting without Verilog code generation.} One may ask: why not let an LLM directly forecast performance and power without generating Verilog code. An experiment is performed to verify this concept and the results from 6 different LLMs are shown in Table~\ref{tab:LLMreasonning}. The Llama 3-8B model has been fine-tuned with the training dataset completely separated from the testcases. 
One can see that their errors are much greater than Lorecast.
For the design ``{\em right\_shift}", Gemini 1.5 Pro could not even produce legal results. The results confirm that direct forecasting using LLMs without Verilog code generation is a significantly more difficult path than Lorecast.

\subsection{Importance of Syntax Correctness and Structural Similarity}
In Table~\ref{tab:generate_comparison}, RTLLM~\cite{lu2024RTLLM} and VerilogEval~\cite{liu2023VerilogEval} are two previous works on LLM-based Verilog code generation. All forecast techniques in Table~\ref{tab:generate_comparison} use the same ML power and TNS prediction
technique~\cite{Sengupta2022Prediction} based on the generated Verilog code. 
There are two observations here. First, achieving a high syntax correctness rate is not straightforward, as both RTLLM~\cite{lu2024RTLLM} and VerilogEval~\cite{liu2023VerilogEval} fail to achieve syntax correctness for a significant number of cases. In contrast, Lorecast achieves $100\%$ syntax correctness in all cases. Second, syntax correctness is critical for the forecast. Without syntax correctness, the ML technique~\cite{Sengupta2022Prediction} is not able to generate power/TNS prediction results.

The results of Table~\ref{tab:generate_comparison} also highlight the importance of structural similarity, which is described in Section~\ref{sec:structural_similarity}. For the two cases ``signal\_generator" and ``pe\_32bit", VerilogEval~\cite{liu2023VerilogEval} produces syntax correct but functionally incorrect Verilog code. This code leads to much greater prediction errors than Lorecast and RTLLM~\cite{lu2024RTLLM} due to their AST structures not being similar to the functionally correct ground truth designs. 
Specifically, for the “signal generator” case, VerilogEval’s AST achieves only 20.8\% Subtree Match Rate\cite{baxter1998clone} with the ground truth, while RTLLM and Lorecast exceed 95\%, with Lorecast also being functionally correct. 
For “pe\_32bit”, both Lorecast and RTLLM produce functionally correct code with match rates exceeding 85\%, whereas VerilogEval remains both functionally incorrect and structurally misaligned, with only 36\% match. These results suggest an association between AST-level structural similarity and the accuracy of performance and power predictions.

\subsection{Impact of Functional Incorrectness}

Comparing the GPT4-based Lorecast and Gemini1.5Pro-based Lorecast in Table~\ref{tab:maintable}, their errors are similar, although Gemini1.5Pro results in 3 functionally incorrect designs while GPT4 has only 2. In general, the impact of functional incorrectness on Lorecast accuracy is small due to the structural similarity described in Section~\ref{sec:structural_similarity}.

We also examine the impact of functionally incorrect Verilog code on logic and layout synthesis outcomes, as shown in Table~\ref{tab:func_incorrect_results}, where power and TNS values are obtained through actual synthesis rather than ML prediction. This experiment is conducted using various LLMs with Lorecast-style prompting. Except for "asyn\_fifo", where some generated Verilog code is not synthesizable, the functionally incorrect Verilog code generally yields synthesis results, both in terms of power and TNS, that closely match those of functionally correct Verilog implementations. This outcome can be attributed to the structural similarity between the Verilog code generated by Lorecast and its functionally correct counterparts, as discussed in Section~\ref{sec:structural_similarity}.

\begin{table}[h]
\vspace{-10pt}
\caption{Effect of functionally incorrect Verilog code generated by Lorecast on logic/layout synthesis (not forecast) results.}
\resizebox{\linewidth}{!}{%
\begin{tabular}{l||cc||cccccccc}
\hline
\multicolumn{1}{c||}{\multirow{3}{*}{\textbf{Design}}} & \multicolumn{2}{c||}{\textbf{Functionally Correct}} & \multicolumn{8}{c}{\textbf{Functionally Incorrect}} \\ \cline{2-11} 
\multicolumn{1}{c||}{} & \multicolumn{2}{c||}{\textbf{Manual}} & \multicolumn{2}{c|}{\textbf{GPT4}} & \multicolumn{2}{c|}{\textbf{Gemini 1.5 Pro}} & \multicolumn{2}{c|}{\textbf{GPT4o}} & \multicolumn{2}{c}{\textbf{DeepSeek V3}} \\
\multicolumn{1}{c||}{} & Power{*} & TNS{*} & Power & \multicolumn{1}{c|}{TNS} & Power & \multicolumn{1}{c|}{TNS} & Power & \multicolumn{1}{c|}{TNS} & Power & TNS \\ \hline
\hline
signal generator & 1508 & 0.225 & $\varnothing$ & \multicolumn{1}{c|}{$\varnothing$} & 1346 & \multicolumn{1}{c|}{0.279} & 1467 & \multicolumn{1}{c|}{0.253} & 1367 & 0.263 \\
asyn\_fifo & 100 & 0 & \ding{55} & \multicolumn{1}{c|}{\ding{55}} & \ding{55} & \multicolumn{1}{c|}{\ding{55}} & 125 & \multicolumn{1}{c|}{0} & \ding{55} & \ding{55} \\
sobel\_filter & 49475 & 0.995 & 48737 & \multicolumn{1}{c|}{0.981} & 50537 & \multicolumn{1}{c|}{1.029} & 47415 & \multicolumn{1}{c|}{1.011} & 46734 & 1.026 \\ \hline
\multicolumn{11}{l}{$\varnothing$: functionally correct and thus out of the scope of this experiment. }\\
\multicolumn{9}{l}{\ding{55}: syntax correct but not synthesizable. }\\
\multicolumn{9}{l}{$^{\mathrm{*}}$Power unit is $\mu W$ and TNS  unit is $ns$.}
\end{tabular}
}
  \label{tab:func_incorrect_results}
  \vspace{-15pt}
\end{table}

\subsection{Ablation Studies}

To assess the general applicability of Lorecast, we additionally conduct ablation studies on the RTLLM benchmark \cite{lu2024RTLLM}, a set of design tasks introduced in prior work, as a complementary evaluation beyond our primary testcases.

\noindent{\bf Impact of different templates in prompting correctness without I-PREF.}
Although template-based structured prompting has been proposed in previous works~\cite{liu2023VerilogEval}~\cite{lu2024RTLLM}, different template styles also matter. In Table~\ref{tab:description_templates}, we show that our template style can improve the syntax correctness rate by $33\% –43\%$. Please note that I-PREF is not performed in these cases. 

\begin{table}[h]
\centering
\vspace{-10pt}
\caption{Impact of different templates on Syntax and Functional Correctness Across LLMs (Without I-PREF).}
\resizebox{0.99\linewidth}{!}{%
\begin{tabular}{|l||cccccccc|}
\hline
\multicolumn{1}{|c||}{\multirow{3}{*}{\textbf{\begin{tabular}[c]{@{}c@{}}Prompt\\ Technology\end{tabular}}}} & \multicolumn{8}{c|}{\textbf{Correctness Rate(\%)}} \\ \cline{2-9} 
\multicolumn{1}{|c||}{} & \multicolumn{2}{c|}{GPT4} & \multicolumn{2}{c|}{GPT4o} & \multicolumn{2}{c|}{Gemini 1.5 Pro} & \multicolumn{2}{c|}{DeepSeek V3} \\
\multicolumn{1}{|c||}{} & Syntax & \multicolumn{1}{c|}{Func} & Syntax & \multicolumn{1}{c|}{Func} & Syntax & \multicolumn{1}{c|}{Func} & Syntax & Func \\ \hline\hline
VerilogEval\cite{liu2023VerilogEval} & 66.7 & \multicolumn{1}{c|}{21.4} & 50 & \multicolumn{1}{c|}{21.4} & 42.9 & \multicolumn{1}{c|}{17.9} & 55.2 & 25 \\
RTLLM\cite{lu2024RTLLM} & 86.2 & \multicolumn{1}{c|}{39.3} & 82.8 & \multicolumn{1}{c|}{42.9} & 71.4 & \multicolumn{1}{c|}{42.9} & 79.3 & 53.6 \\
Lorecast & 100 & \multicolumn{1}{c|}{57} & 93.1 & \multicolumn{1}{c|}{50} & 86.2 & \multicolumn{1}{c|}{57} & 93.1 & 57 \\ \hline
\end{tabular}%
}
  \vspace{-5pt}
  \label{tab:description_templates}%
\end{table}

\begin{figure}[h] 
\centering 
\vspace{-5pt}
\includegraphics[width=0.95\linewidth]{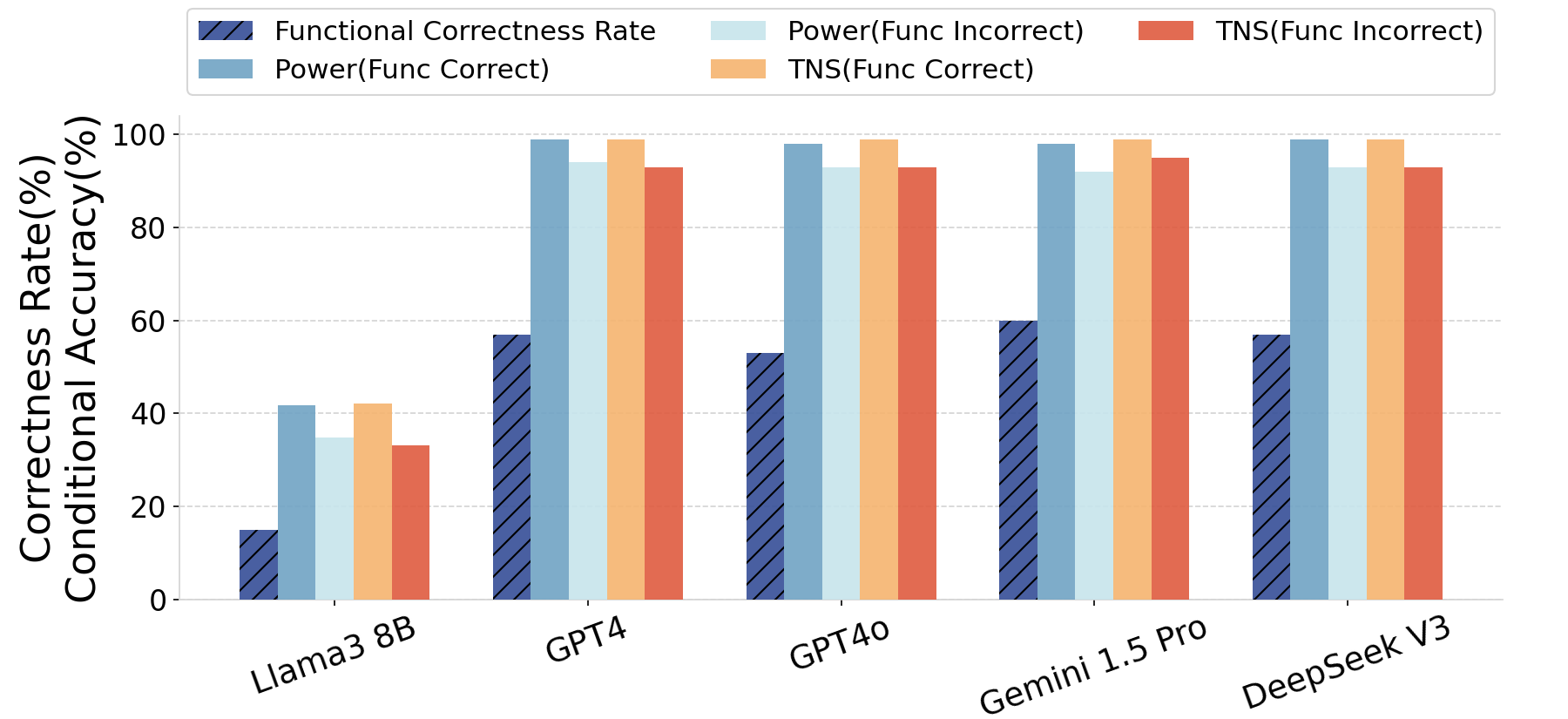}
\vspace{-5pt}
\caption{Comparison of LLMs on forecasting accuracy and functional correctness.}
\vspace{-5pt}
\label{fig:correctness2predict}
\end{figure}

\noindent
{\bf Effect of functional correctness across LLMs.} 
Sometimes, an LLM cannot produce syntax correct Verilog code for a design.
As shown in Figure~\ref{fig:correctness2predict}, the results indicate a correlation between functional correctness(represented by the hatched bars) and conditional accuracy, although functional errors can be tolerated. 
For Lorecast@GPT4, power and performance $\mathcal{E}$ is 1\% for functionally correct code, rising to 5\% and 7\% for functionally incorrect one.

\begin{figure}[h] 
    \centering
    \includegraphics[width=0.68\linewidth]{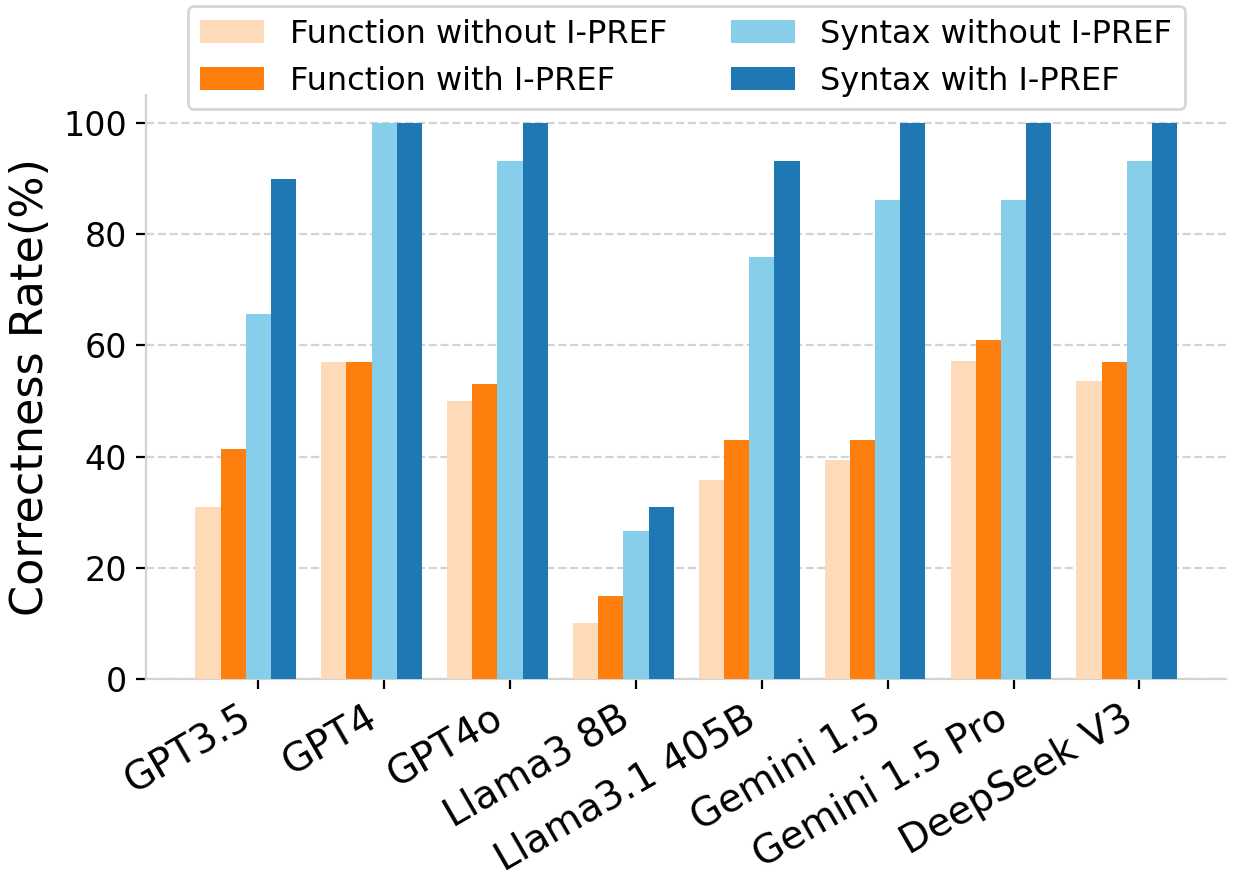}
    \vspace{-3pt}
    \caption{Syntax/functional correctness of LLMs with and without I-PREF.}
    \label{fig:Self-correction}
    \vspace{-12pt}
\end{figure}

\noindent
{\bf Impact of I-PREF on correctness across LLMs. }
Syntax/functional correctness rates for 8 LLMs with and without I-PREF are depicted in Figure~\ref{fig:Self-correction}. Here, syntax/functional correctness is asserted for a design if the results from all three attempts are correct.

We can obtain the following observations.

\begin{itemize}[nosep]
    \item Syntax correctness rate is always significantly higher than functional correctness rate. This confirms that syntax correctness is a much more achievable goal than functional correctness for LLM-generated Verilog code.
    \item I-PREF can always improve syntax correctness rate, which is fundamental for Lorecast.
\end{itemize}

\begin{figure}[h] 
    \centering
    \vspace{-12pt}
    \includegraphics[width=0.68\linewidth]{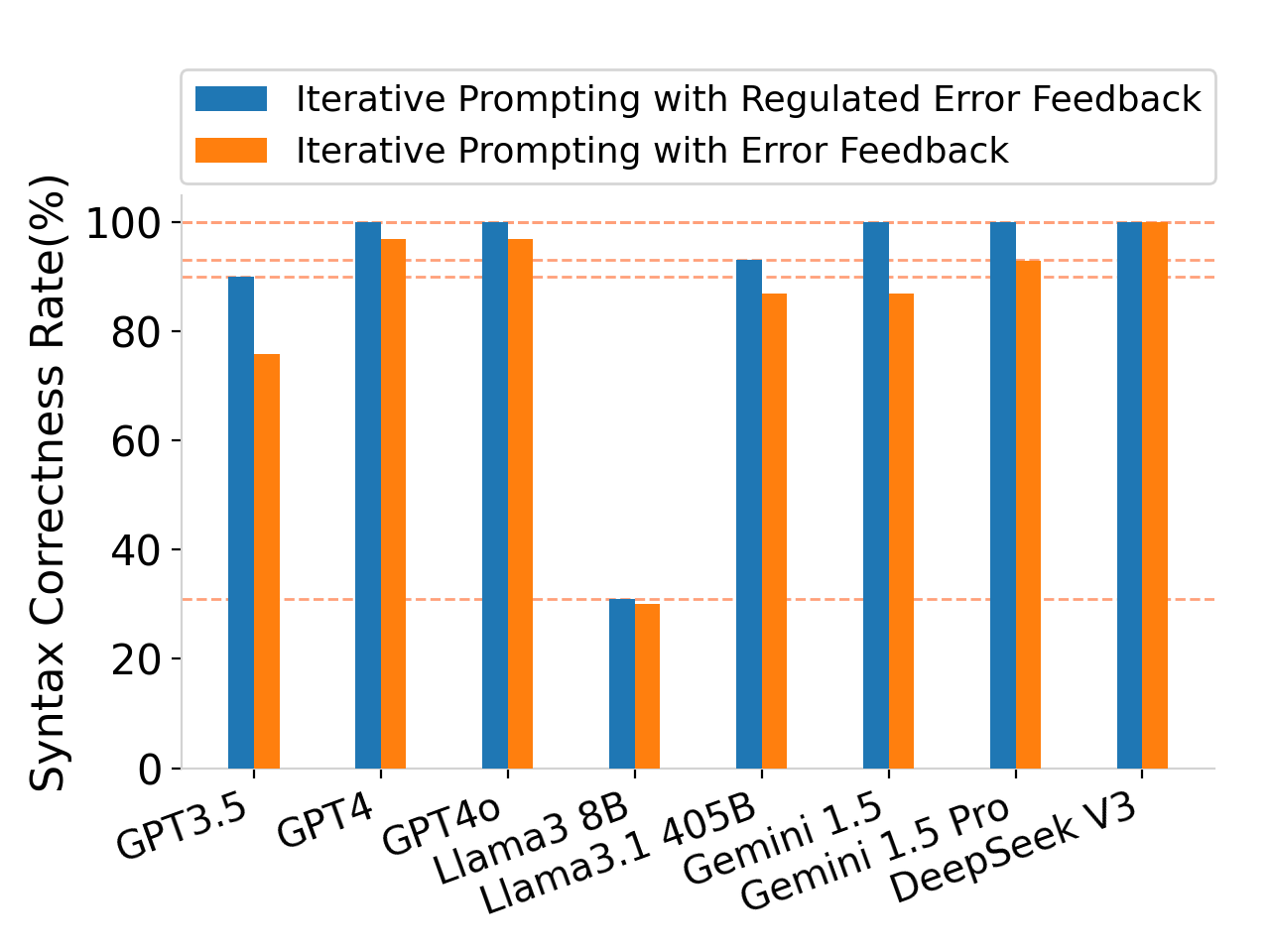}
    \vspace{-10pt}
    \caption{Effect of regulation in iterative feedback prompting.}
    \label{fig:Regulated_correct}
    \vspace{-4pt}
\end{figure}

\noindent
{\bf Effect of regulation in iterative feedback prompting on correctness.}
Although iterative feedback prompting was proposed in \cite{Thakur2024AutoChip}, its feedback prompting is not regulated. Figure~\ref{fig:Regulated_correct} shows that our regulated feedback prompting in Lorecast can improve the syntax correctness rate by about 6\%.


\begin{figure}[h] 
    \centering
    \vspace{-4pt}
    \includegraphics[width=0.65\linewidth]{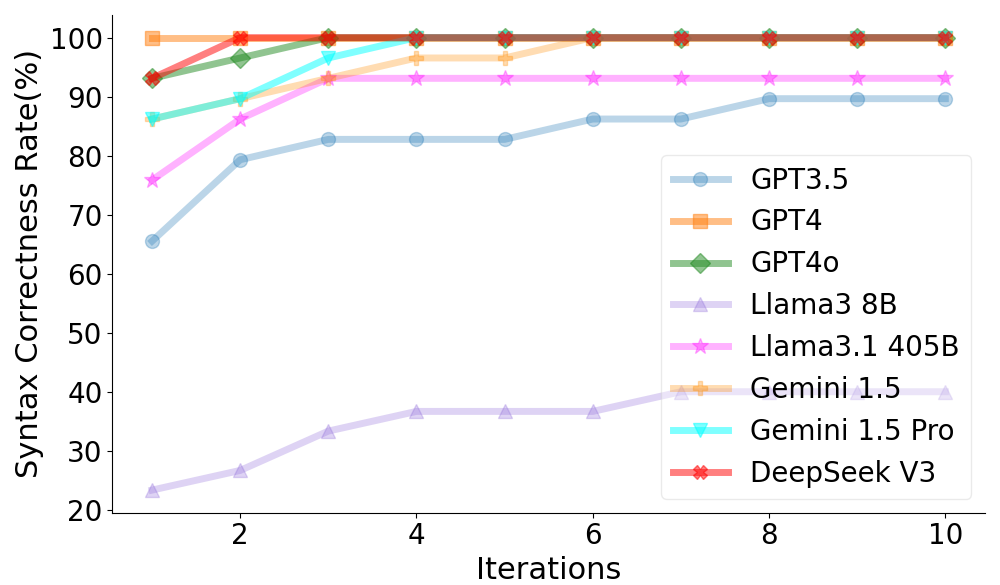}
    \vspace{-5pt}
    \caption{Syntax correctness rate vs. the max I-PREF iterations.}
    \label{fig:different_iteration}
    \vspace{-10pt}
\end{figure}

\noindent{\bf Syntax correctness rate versus max I-PREF iterations.} In Figure~\ref{fig:different_iteration}, we vary the maximum number of I-PREF iterations and observe the impact on syntax correctness rate. Initially, increasing the maximum number of iterations does help in improving the correctness rate. However, the benefits diminish at around 8 iterations. This is why we set the limit of I-PREF iterations to 10.


\section{Conclusions}
\label{sec:conclusions}

We propose a methodology for forecasting circuit performance and power from natural language. It leverages LLM-based Verilog code generation techniques and the Verilog-based ML prediction technique to produce forecasts with approximately $2\%$ error compared to post-layout analysis. Lorecast accelerates the performance and power estimation process by 4.6× compared to conventional methods that involve manually writing Verilog code, logic and layout synthesis. The LLM-based Verilog code generation technique here is customized to improve syntax correctness rate and reduce the requirement for functional correctness. The validation is performed on circuits significantly larger than previous works on LLM-based Verilog code generation.

\newpage
\bibliographystyle{IEEEtran}
\bibliography{ref.bib}
\end{document}